\documentclass[10pt,twocolumn]{article}
\usepackage[pdftex]{graphicx}
\usepackage{amsmath,amssymb}
\usepackage[T1]{fontenc}
\usepackage[english]{babel}
\usepackage[colorlinks,linkcolor=blue,urlcolor=blue,citecolor=black]{hyperref}
\usepackage{mathptmx}
\usepackage{rotating}
\usepackage{fancyhdr}
\usepackage{caption}
\usepackage[super]{natbib}
\bibpunct{}{}{,}{s}{}{\textsuperscript{,}}

\usepackage{geometry}
\newcommand*\aap{A\&A}

\newcommand*\aaps{A\&AS}

\newcommand*\aj{AJ}

\newcommand*\apj{ApJ}
\newcommand*\apjl{ApJ}

\newcommand*\apjs{ApJS}

\newcommand*\araa{ARA\&A}

\newcommand*\mnras{MNRAS}

\newcommand*\nat{Nature}

\newcommand*\pasp{PASP}

\begin{document}
\small
\title{$\gamma$\,Columbae: the recently stripped, pulsating core of a massive star}
\author{Andreas Irrgang${}^1$*, Norbert Przybilla${}^2$, Georges Meynet${}^3$}
\date{\vspace{-3ex}}
\maketitle
\thispagestyle{fancy}
\renewcommand{\headrulewidth}{0pt}
\renewcommand{\footrulewidth}{0.4pt}
\fancyhf{}
\fancyhead[L]{\color{red}{\footnotesize This preprint has not undergone peer review or any post-submission improvements or corrections. The Version of Record of this
article is published in Nature Astronomy, and is available online at \href{https://doi.org/10.1038/s41550-022-01809-6}{https://doi.org/10.1038/s41550-022-01809-6}.}}
\fancyfoot[L]{\footnotesize ${}^1$Dr.~Karl~Remeis-Observatory \& ECAP, Friedrich-Alexander University Erlangen-Nuremberg, Bamberg, Germany. ${}^2$Institut f\"ur Astro- und Teilchenphysik, Universit\"at Innsbruck, Innsbruck, Austria. ${}^3$Geneva Observatory, University of Geneva, Sauverny, Switzerland. *e-mail: \href{mailto:andreas.irrgang@fau.de}{andreas.irrgang@fau.de}}
{\bf\noindent
A vital condition for life on Earth is the steady supply of radiative heat by the Sun. Like all other stars, the Sun generates its emitted energy in its central regions where densities and temperatures are high enough for nuclear fusion processes to take place. Because stellar cores are usually covered by an opaque envelope, most of our knowledge about them and their life-giving nuclear processes comes from theoretical modelling\cite{1958ses..book.....S} or from indirect observations such as the detection of solar neutrinos\cite{1968PhRvL..20.1205D} and the study of stellar pulsations\cite{1976Natur.259...89C,2010aste.book.....A}, respectively. Only in very rare cases, stars may expose their cores, e.g., when a tiny fraction of them evolves into Wolf-Rayet\cite{2007ARA&A..45..177C} or helium hot subdwarf\cite{2016PASP..128h2001H} stars. However, for the vast majority of stars, namely unevolved stars that burn hydrogen to helium in their centres, direct observational clues on the cores are still missing. Based on a comprehensive spectroscopic and asteroseismic analysis, we show here that the bright B-type star $\gamma$\,Columbae is the stripped pulsating core (with a mass of $4$--$5\,M_\odot$, where $M_\odot$ is the mass of the Sun) of a previously much more massive star of roughly $12\,M_\odot$ that just finished central hydrogen fusion. The star's inferred parameters indicate that it is still in a short-lived post-stripping structural readjustment phase, making it an extremely rare object. The discovery of this unique star paves the way to obtain invaluable insights into the physics of both single and binary stars with respect to nuclear astrophysics and common-envelope evolution. In particular, it provides first observational constraints on the structure and evolution of stripped envelope stars\cite{2018A&A...615A..78G,2019ApJ...878...49W}.
}

\begin{figure}[t]
\centering
\includegraphics[width=0.49\textwidth]{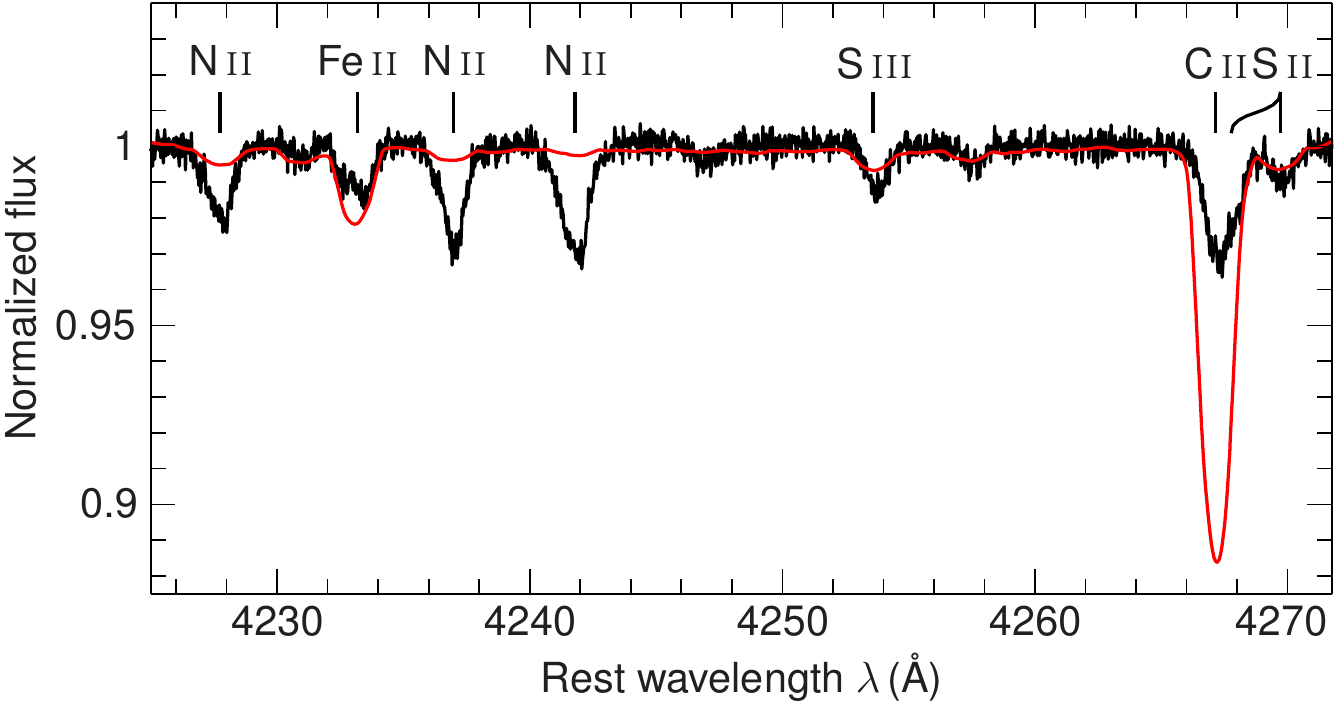}
\caption{\small {\bf Observational signature of the peculiar abundances of carbon and nitrogen.} A small portion of an observed spectrum of $\gamma$\,Columbae (black line) is compared to one of the solar-neighbourhood reference star 18\,Pegasi (red line; smoothed in such a way that the broadening of the spectral lines is comparable), which has similar atmospheric parameters ($T_{\mathrm{eff}}=15,800$\,K, $\log(g\,\mathrm{(cm\,s^{-2})})=3.75$) but a standard, solar-like chemical composition\cite{2012A&A...539A.143N}. Singly and doubly ionized lines of nitrogen, iron, sulphur, and carbon are labelled. Already a visual inspection of the two spectra shows that the abundances, i.e., the strengths of the absorption lines, of iron and sulphur (and of other chemical species whose spectral lines are not exhibited here) are very similar while carbon is severely depleted and nitrogen considerably enriched in the atmosphere of $\gamma$\,Columbae.
}
\label{fig:spectra_zoom}
\end{figure}

The star $\gamma$\,Columbae is one of the brightest stars in the constellation Columba and is thus one of only a few thousands out of the many billions of stars in our Milky Way that we can observe directly with the naked eye. Intuitively, one would think that such a prominent star has been well studied in the past and is now well understood. Rather the opposite is true, however. Although high-quality data, i.e., spectra with high resolving power and high signal-to-noise ratio, of this target have been taken, there are no reports in the literature on their analysis with respect to the determination of stellar parameters and chemical elemental abundances. Those spectra were solely used to classify the object as a candidate slowly pulsating star of spectral type B based on distortions in the line profiles\cite{2006A&A...452..945T} or to perform magnetic field measurements, which, however, led to contradictory results\cite{2009AN....330..317H,2012A&A...538A.129B}. Only the quantitative spectral analysis that is presented here reveals that $\gamma$\,Columbae, which at first glance looks like a typical star in the solar neighbourhood, is anything else but normal. While the atmospheric parameters (effective temperature $T_{\mathrm{eff}} = 15,570 \pm 320$\,K and surface gravity $\log(g\,\mathrm{(cm\,s^{-2})}) = 3.3 \pm 0.1$; 99\% confidence intervals) indicate that the star could just be an ordinary subgiant of spectral class B, a standard evolutionary history can be excluded right away on the basis of the derived surface abundances of helium, carbon, and nitrogen, which, unlike those of several heavier chemical elements, are completely different from normal stars (Fig.~\ref{fig:spectra_zoom}).%
Instead, this peculiar abundance pattern is characteristic of the carbon-nitrogen-oxygen (CNO) cycle, which is the energy generation mechanism at the cores of stars that are more massive than the Sun. Those stars fuse hydrogen to helium via reaction cycles that involve isotopes of carbon, nitrogen, and oxygen as catalysts\cite{1938PZ....39...633v,1939PhRv...55..434B}. Typical qualitative signatures of the CNO cycle are increased abundances of helium and nitrogen at the expense of decreased abundances of hydrogen, carbon, and oxygen. Because those elements' quantitative abundance ratios are very sensitive to the properties of the stellar plasma, in particular to its temperature, they are excellent probes for the conditions inside the hydrogen burning regions, i.e., for the cores of stars. 

\begin{figure}[t]
\centering
\includegraphics[width=0.49\textwidth]{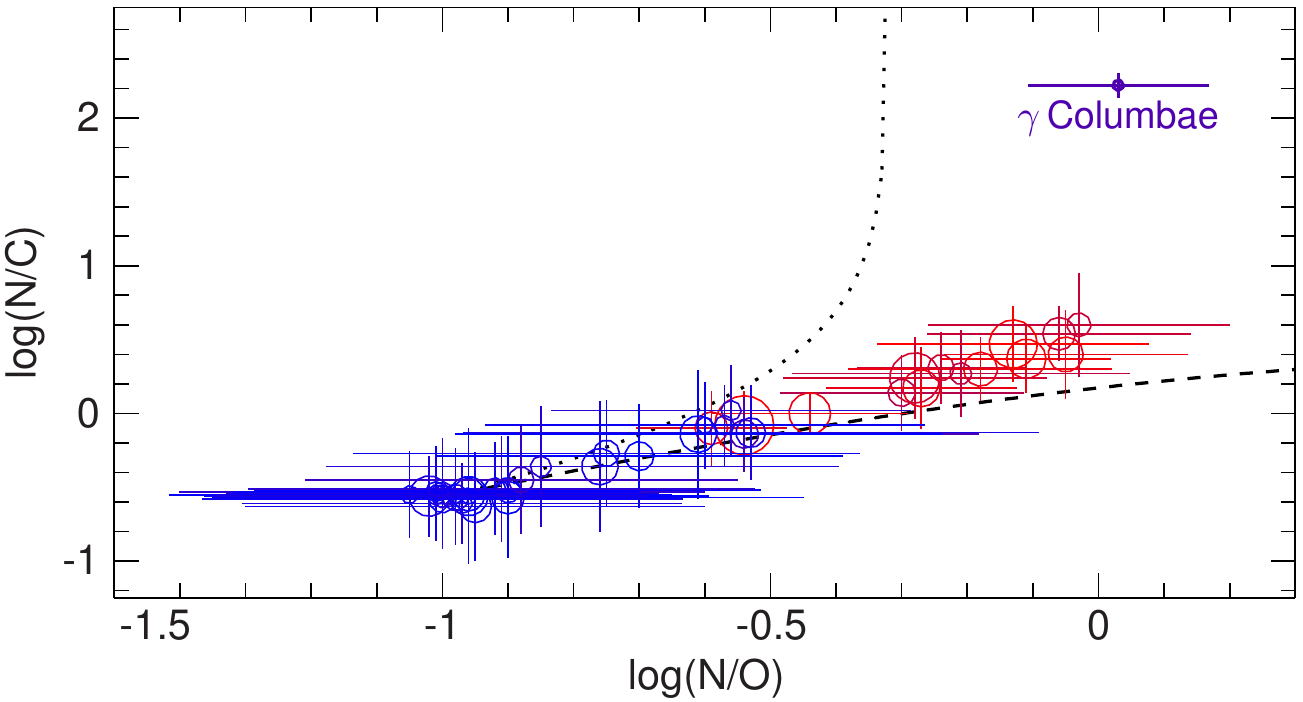}
\caption{\small {\bf Comparison with CNO signatures from other stars.}  Number ratios of nitrogen over carbon versus nitrogen over oxygen are plotted for a sample of 45 reference stars\cite{2012A&A...539A.143N,2010A&A...517A..38P,2011A&A...532A...2N,2015A&A...574A..20F} that cover a wide range of masses (6 to 27\,$M_\odot$; the mass scales with the size of the circles) and evolutionary stages ($1.2 \lesssim \log(g\,\mathrm{(cm\,s^{-2})}) \lesssim 4.3$; color coded: the bluer the symbol the larger the surface gravity and the less evolved the star). The black lines are analytic predictions for the CNO cycle based on simplified assumptions\cite{2014A&A...565A..39M} (\textit{dashed:} ON-cycle limit, carbon remains constant at low CN equilibrium value; \textit{dotted:} CN-cycle limit, oxygen remains constant at its initial abundance). Unlike stars of similar mass and surface gravity, which group at the bottom left corner of this diagram where abundances are close to their pristine values, $\gamma$\,Columbae lies isolated at the top right corner, demonstrating that its surface CNO pattern is too extreme to be explained by standard stellar evolution, making it unique. Error bars represent 99\% confidence intervals.
}
\label{fig:CNO}
\end{figure}

Although the nuclear ashes usually remain hidden in the stellar interiors, there are a few mechanisms which can make them observable. Hydrodynamic mixing of core fusion products with pristine material in the stellar envelope can lead to a modest but still detectable footprint of the CNO cycle in the emitted spectra\cite{2010A&A...517A..38P,2014A&A...565A..39M,2015A&A...575A..34M}. However, the derived abundance pattern for $\gamma$\,Columbae with its strong CN-cycle signature -- the number ratio of nitrogen to oxygen atoms, which is in the following abbreviated by N/O, is comparable to heavily mixed evolved stars while N/C is about 50 times larger -- is far too extreme to be explained by this process (Fig.~\ref{fig:CNO}).%
For the same reasoning, accretion of CNO processed material from a companion can be excluded as well. Two classes of massive stars that show highly processed CNO signatures are known for a long time. The ON stars\cite{1971ApJ...164L..67W} follow the trend indicated by the comparison stars in Fig.~\ref{fig:CNO} to higher values\cite{2015A&A...578A.109M}, reaching maximum values among the Wolf-Rayet stars of WN subtype at about $\log(\textnormal{N/C})\approx2$ at $\log(\textnormal{N/O})\approx1.6$\cite{2001ApJ...548..932H}. While the Wolf-Rayet stars represent the late evolutionary stage of initially very massive stars whose powerful winds are able to strip off their envelopes and expose their CNO processed cores, the precise mechanism for CNO enrichment for the ON stars is still unclear\cite{2015A&A...578A.109M}. Because the members of those two classes are much more massive, hotter ($T_{\mathrm{eff}}\gtrsim30,000$\,K), and show a full CNO-cycle signature, a common origin with $\gamma$\,Columbae is highly unlikely.

\begin{figure}[t]
\centering
\includegraphics[width=0.49\textwidth]{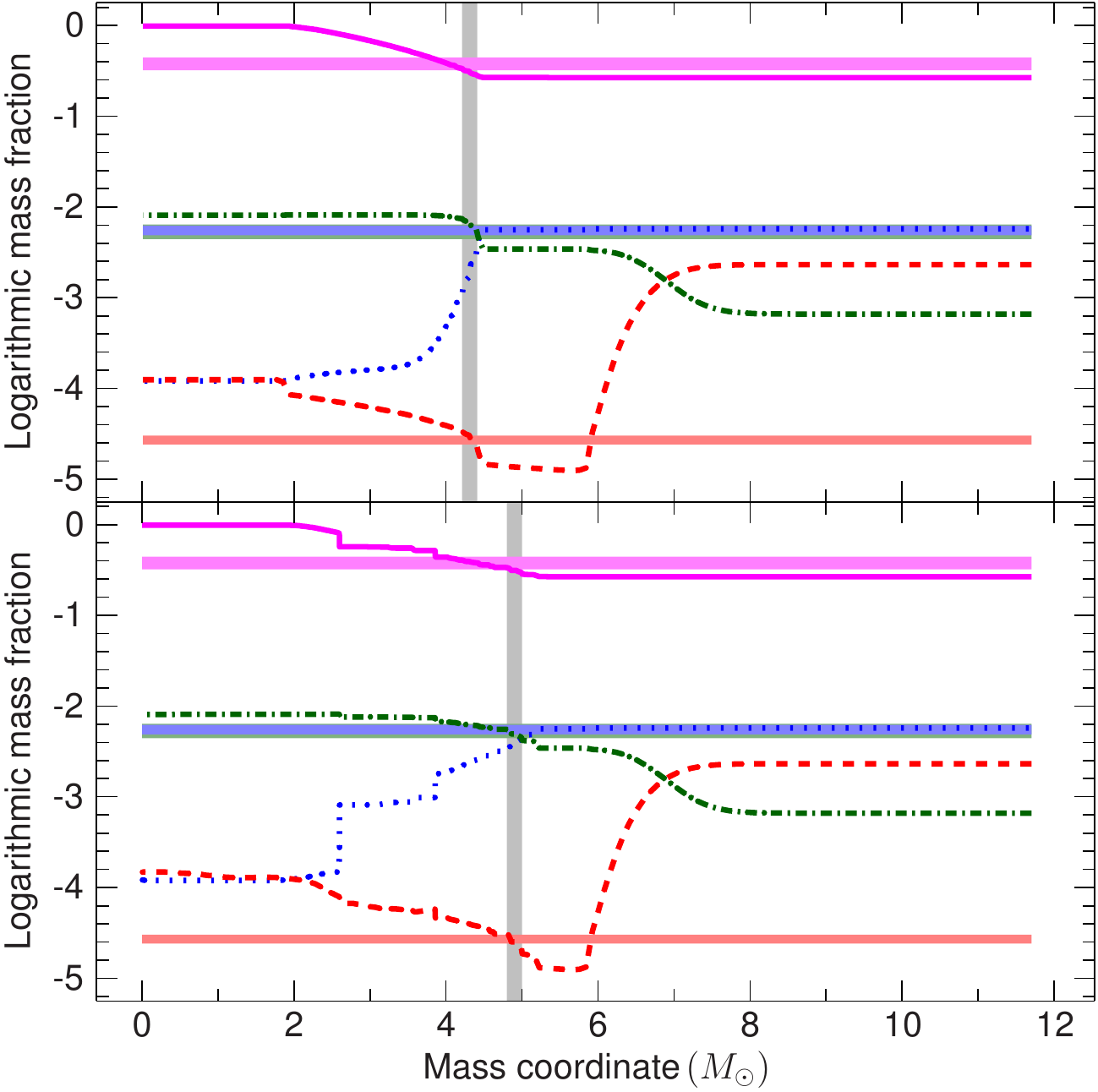}
\caption{\small {\bf Comparison between observed abundances and theoretical predictions for the depth-dependent chemical composition in a massive star.} Logarithmic mass fractions for helium (magenta solid line), carbon (red dashed line), nitrogen (green dashed-dotted line), and oxygen (blue dotted line) are plotted as a function of interior mass for a non-rotating stellar model\cite{2012A&A...537A.146E} with solar metallicity and initial mass of $12\,M_\odot$ at two moments shortly after the end of central hydrogen fusion (\textit{top}: age $\tau=15.48$\,Myr, radius $R_\star=8.6\,R_\odot$; \textit{bottom}: $\tau = 15.53$\,Myr, $R_\star=16.1\,R_\odot$). The observed surface abundances of $\gamma$\,Columbae, whose 99\% confidence intervals are represented by coloured horizontal bars, are well reproduced by the model for interior masses marked by gray vertical bars, which indicates that the object could be the stripped $4$--$5\,M_\odot$ core of a $12\,M_\odot$ star. The mixture of those chemical elements is not only a very sensitive function of interior mass and stellar age, but also of the total mass.}
\label{fig:Geneva_stellar_structure}
\end{figure}

Finally, stars may also get stripped via mass transfer in a binary system. The class of helium hot subdwarf stars\cite{2016PASP..128h2001H} is suggested to be the result of such a binary interaction that turns a low-mass donor star into a hot ($T_{\mathrm{eff}}\gtrsim35,000$\,K) and compact ($\log(g\,\mathrm{(cm\,s^{-2})})\gtrsim5$) object of roughly 0.5\,$M_\odot$ whose surface abundances show heavy imprints of the CNO cycle. Although $\gamma$\,Columbae is clearly different from a hot subdwarf star in terms of atmospheric parameters, it might still be the stripped star in a binary system, except that the low-mass donor star is replaced by a more massive one. The motivation for this is that the observed mass fractions of those chemical elements that are affected by the CNO cycle nicely match the predicted abundances of a $4$--$5\,M_\odot$ core of a $12\,M_\odot$ stellar model close to the end of central hydrogen fusion (Fig.~\ref{fig:Geneva_stellar_structure}).%
This idea is corroborated by noting that such a core mass is consistent with independent estimates of the star's current mass based on the observed spectral line-profile variations ($4.0\,M_\odot$) and on its measured parallactic distance ($5.0\pm1.7\,M_\odot$; 99\% confidence interval).

According to stellar evolution theory, exhaustion of central hydrogen fusion is accompanied by an expansion of the star. If occurring in a binary system, this expansion can lead to an exchange of mass between the two components, which is dynamically stable for mass ratios that are not too extreme, e.g., for masses of $12\,M_\odot$ for the primary and $5\,M_\odot$ for the secondary component, respectively. A tailored binary evolution model for such a system\cite{2017A&A...608A..11G} shows that binary stripping can indeed lead to an exposed core of $\sim4\,M_\odot$, which, however, would then be the less luminous component in the binary system. Because there are no indications for a luminous companion close to $\gamma$\,Columbae, this scenario cannot be valid here. However, if the mass of the companion is $\lesssim3\,M_\odot$, mass transfer becomes unstable\cite{2018A&A...615A..78G}, resulting in a so-called common-envelope phase, in which both components orbit each other within the extended envelope of the donor. Although the detailed physics of this complex phase cannot be precisely modelled yet, conservation of momentum and energy dictates that the two components spiral in and that the common envelope is eventually expelled from the system\cite{2011ApJ...730...76I}, leaving behind a stripped core whose low-mass companion is faint enough to go unnoticed next to it. Although this scenario is not unlikely to occur given that extreme mass ratios are quite common for binary systems that contain massive stars\cite{2012Sci...337..444S}, no such system has been reported yet. Apart from some Wolf-Rayet plus O-type star systems\cite{2007ARA&A..45..177C}, only one massive (non-subdwarf) stripped star near its end stage is known\cite{2008A&A...485..245G}.

\begin{figure}[t]
\centering
\includegraphics[width=0.49\textwidth]{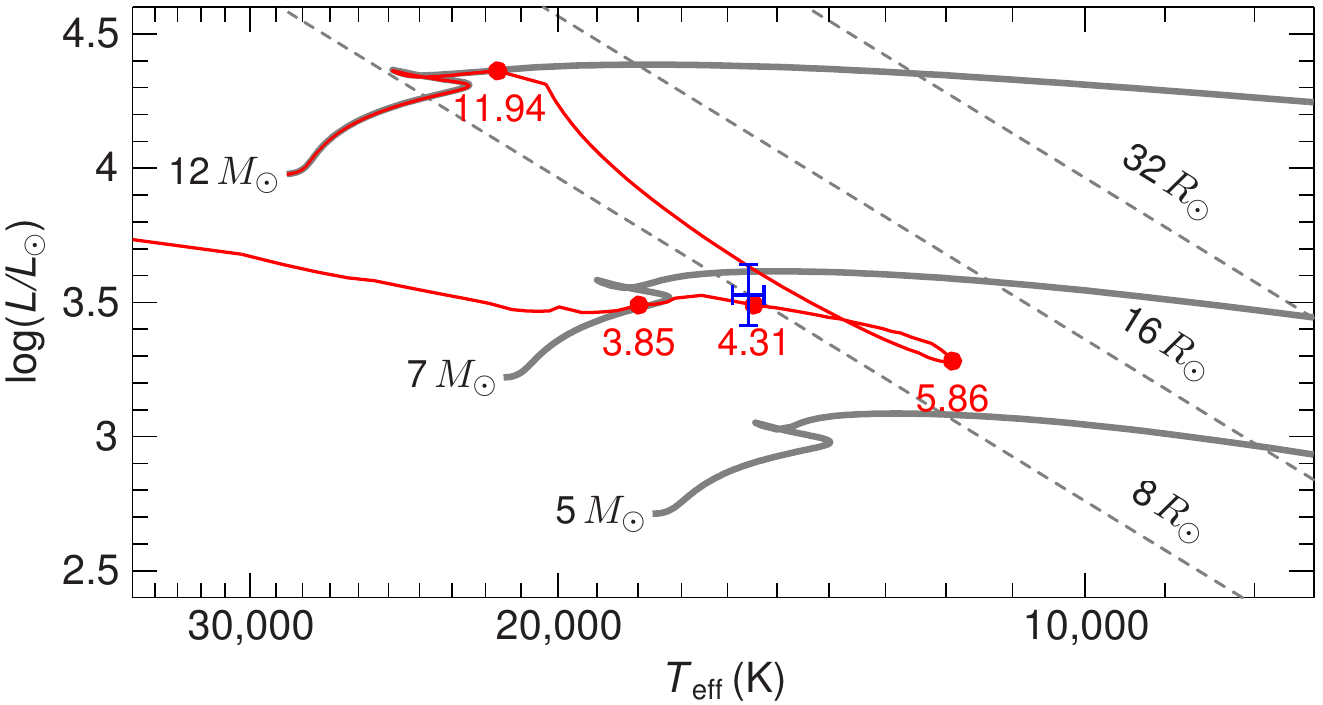}
\caption{\small {\bf Stellar evolution in the Hertzsprung-Russell diagram.} The evolution of effective temperature $T_{\mathrm{eff}}$ and stellar luminosity $L$ for a non-rotating stellar model with solar metallicity and initial mass of $12\,M_\odot$ --~with an additional artificial mass loss of $1.29\times10^{-3}\,M_\odot$\,yr$^{-1}$ switched on shortly after the end of central hydrogen fusion~-- is shown (red solid line; the numbers next to the points indicate the mass in $M_\odot$ at that point in time). The position of $\gamma$\,Columbae (blue; error bars are 99\% confidence intervals), evolutionary tracks\cite{2012A&A...537A.146E} for non-rotating single stars of solar metallicity (gray solid lines), and loci of constant radius (gray dashed lines) are shown for reference. The stripped model with a respective mass of $4.31\,M_\odot$ exhibits surface abundances as indicated by the gray vertical bar in the top panel of Fig.~\ref{fig:Geneva_stellar_structure} and is thus not only able to reproduce $\gamma$\,Columbae's stellar parameters (mass, radius, and luminosity) but also its observed mass fractions for helium, carbon, nitrogen, and oxygen.}
\label{fig:evolution_tracks}
\end{figure}

To see whether $\gamma$\,Columbae's observed stellar parameters are consistent with this scenario, we followed the evolution of a $12\,M_\odot$ stellar model under the assumption that the ejection of the common envelope can be mimicked by the evolution of a single star for which an artificial mass loss is switched on shortly after the depletion of central hydrogen fusion\footnote{We note that the star does not experience such a strong mass loss at present.}. Although tailored binary evolution models with a more realistic treatment of common-envelope ejection are desirable in the end, we consider this procedure to be realistic enough to illustrate the effects of sudden mass loss. The result of this exercise is very promising and in qualitative agreement with more sophisticated binary evolution models\cite{2018A&A...615A..78G}. Before ending as a hot and compact object, which is the typical fate of stripped stars, the stellar model becomes significantly cooler and less luminous as a consequence of structural readjustments triggered by the sudden drop in mass, passing the currently observed position of $\gamma$\,Columbae in the Hertzsprung-Russell diagram (Fig.~\ref{fig:evolution_tracks}).%
The timescale of this dynamical and thermal restructuring is of the order of ten thousand years, which is just one thousandth of the total lifetime of a $12\,M_\odot$ star. Because of the rareness of massive stars\cite{2018A&A...615A..78G}, one would expect to find only very few objects in this extremely short-lived evolutionary phase among the known massive stars in the Milky Way. This renders the bright $\gamma$\,Columbae a unique testbed for stellar (binary) evolution, so far hidden in plain sight.

Dynamical friction during the common-envelope phase would lead to a shrinkage of the orbital separation and, hence, to a compact binary system with relatively large velocity semiamplitudes. As a consequence of the increased tidal forces, the two components could also be spun up over time. Possible observational signatures for a common-envelope scenario could therefore include radial velocity variations and pronounced stellar rotation, none of which is however observed in the case of $\gamma$\,Columbae. The reason for this could simply be that the binary system is seen relatively pole-on. Intuitively, one might also expect that the ejection of 7--8\,$M_\odot$ of stellar envelope less than a few thousand years ago leaves behind an easily detectable observational signature in form of, e.g., infrared dust emission. Yet, no hint of circumstellar material is found in mid-infrared images\footnote{The ejection of a common envelope occurs at about the escape velocity from the object's surface, which implies that the ejected material could nowadays extent over a few parsecs, which translates to tens of arcminutes diameter at the star's distance.} taken by the Wide-field Infrared Survey Explorer mission\cite{2010AJ....140.1868W}. The chemical composition of the ejecta is of pristine composition, i.e., mostly hydrogen. During the expansion of the envelope, the material reaches low temperatures at relatively high densities, such that three-body reactions can efficiently transform the atomic hydrogen almost completely into H$_2$ molecules, like in circumstellar envelopes\cite{1983MNRAS.203..517G}. Molecular hydrogen (and helium) is difficult to detect, which is why the bulk of the ejecta could remain invisible in most wavelengths because $\gamma$\,Columbae is yet not hot enough to photoionize and excite this material. High-quality observations in the far ultraviolet, which are unavailable for the star, would be required to trace the H$_2$ Lyman band transitions. Future investigations in this direction could provide valuable insights into the details of common-envelope ejection, a hydrodynamic process that is poorly understood so far.

A better understanding for the inner structure of $\gamma$\,Columbae can come from asteroseismology. Our quantitative modelling of the observed variations in the light curve as well as in the spectral line profiles suggest that $\gamma$\,Columbae is not only a massive star's exposed core that is currently in its post-stripping readjustment phase. It also shows the characteristics of a slowly pulsating B star with an unprecedentedly long observed oscillation period of several years, which is most likely caused by a chance interplay of rotation and oscillation frequency. Future asteroseismic analyses of this unique object based on better light curves and time-series spectroscopy will put first observational constraints on the structure of this long-sought class of massive stripped-envelope stars\cite{2018A&A...615A..78G} and their evolution\cite{2019ApJ...878...49W}.

\begingroup
\renewcommand{\section}[2]{}
\setlength{\bibsep}{0pt}

\endgroup

{\textbf{Acknowledgements}
GM is thankful to the Swiss National Science Foundation (project number 200020-172505).
We thank Ulrich Heber and Manami Sasaki for helpful discussions and John E.\ Davis for the development of the {\sc slxfig} module which was used to prepare the figures in this paper.
Based on data products from observations made with ESO Telescopes at the La Silla Paranal Observatory under program IDs 080.D-0333(A), 182.D-0356(C), 088.A-9003(A), 090.D-0358(A).
Based on observations obtained at the Canada-France-Hawaii Telescope (CFHT) which is operated by the National Research Council of Canada, the Institut National des Sciences de l'Univers of the Centre National de la Recherche Scientique of France, and the University of Hawaii.
Some of the data presented in this paper were obtained from the Mikulski Archive for Space Telescopes (MAST). STScI is operated by the Association of Universities for Research in Astronomy, Inc., under NASA contract NAS5-26555.
This work has made use of data from the European Space Agency (ESA) mission {\it Gaia} (\url{https://www.cosmos.esa.int/gaia}), processed by the {\it Gaia} Data Processing and Analysis Consortium (DPAC, \url{https://www.cosmos.esa.int/web/gaia/dpac/consortium}). Funding for the DPAC has been provided by national institutions, in particular the institutions participating in the {\it Gaia} Multilateral Agreement.
This publication makes use of data products from the Two Micron All Sky Survey, which is a joint project of the University of Massachusetts and the Infrared Processing and Analysis Center/California Institute of Technology, funded by the National Aeronautics and Space Administration and the National Science Foundation.
This publication makes use of data products from the Wide-field Infrared Survey Explorer, which is a joint project of the University of California, Los Angeles, and the Jet Propulsion Laboratory/California Institute of Technology, funded by the National Aeronautics and Space Administration.
}
\clearpage
\section*{Methods}
\section{Photometric analysis}\label{sect:photometry}
\subsection{Spectral energy distribution}
\begin{figure*}[h]
\centering
\includegraphics[width=1\textwidth]{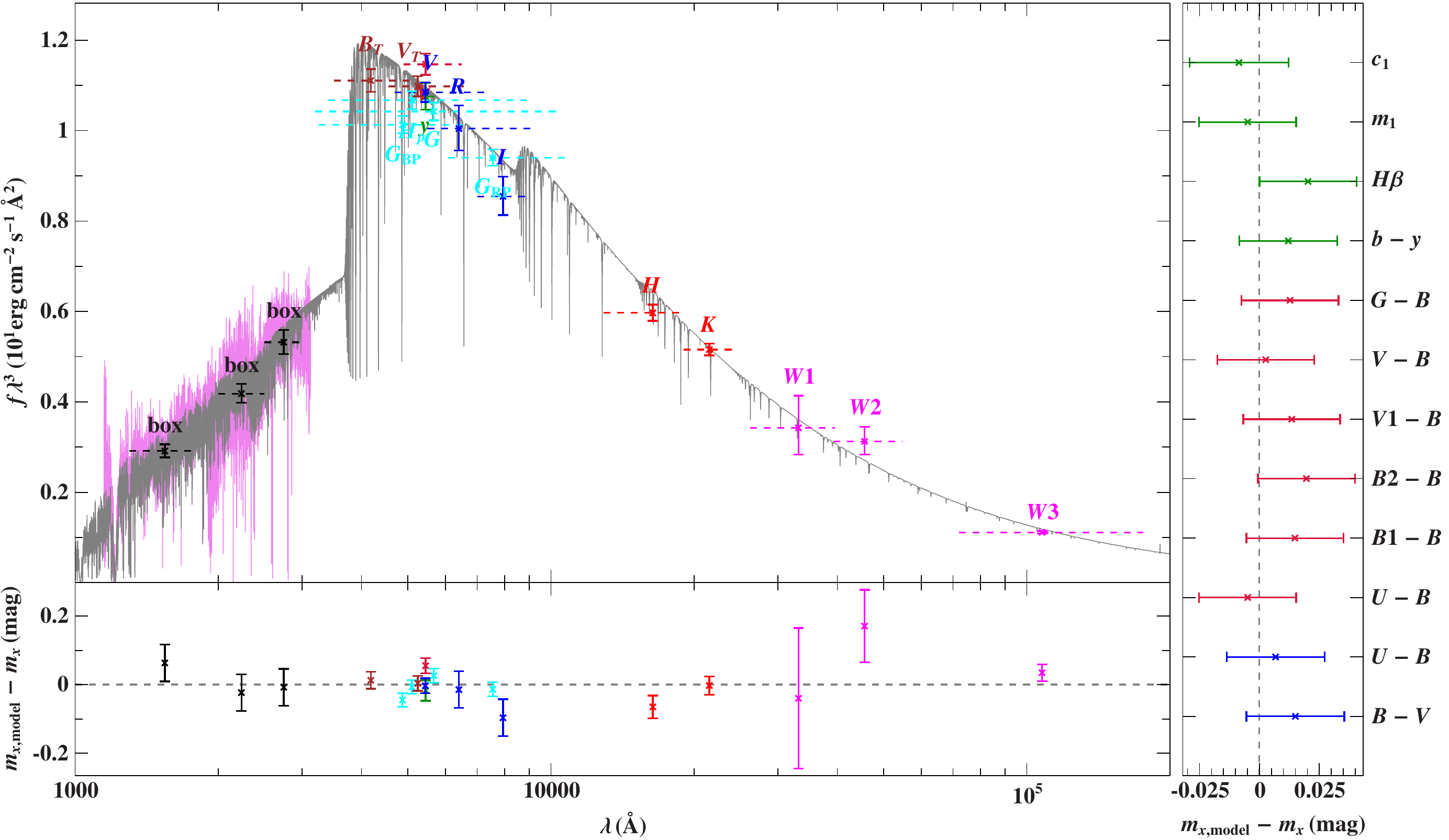}
\caption{\small Comparison of synthetic and observed photometry: The \textit{top panel} shows the spectral energy distribution. The coloured data points are filter-averaged fluxes which were converted from observed magnitudes (the respective filter widths are indicated by the dashed horizontal lines), while the gray solid line represents the best-fitting model (degraded to a spectral resolution of 0.2\,\AA). The three black data points labelled ``box'' are fluxes converted from magnitudes computed by means of box filters of the indicated width from IUE spectra (magenta line). The residual panels at the \textit{bottom} and on the \textit{side} show the differences between synthetic and observed magnitudes and colours, respectively. The photometric systems have the following colour code: Tycho (brown); {\it Gaia} and {\sc Hipparcos} (cyan); Johnson-Cousins (blue); Str\"omgren (green); Geneva (crimson); 2MASS (red); WISE (magenta).}
\label{fig:photometry}
\end{figure*}

The analysis of the spectral energy distribution (SED) is based on photometric data compiled from various catalogues in the literature\cite{1978A&AS...34..477M,1999yCat.2169....0R,2000A&A...355L..27H,2006yCat.2168....0M,2007A&A...474..653V,2014yCat.2328....0C,2015A&A...580A..23P,2018A&A...616A...1G}. Combined with high-dispersion spectra from the International Ultraviolet Explorer (IUE), which are publicly accessible in the MAST archive\footnote{\url{http://archive.stsci.edu/}} under data IDs SWP53044 and LWP29687, the available measurements cover the ultraviolet, optical, and infrared regime. To account for systematic uncertainties such as the intrinsic variability of the star (see Sect.~\ref{subsection:light_curve}), a generic uncertainty of $0.02$\,mag is added in quadrature to all given uncertainties. An improved version of the {\sc Atlas12} code\cite{1996ASPC..108..160K,2018A&A...615L...5I} including level dissolution for hydrogen and non-LTE corrections is employed to compute synthetic SEDs, which are then fitted to the observations by varying the effective temperature $T_\textnormal{eff}$, the surface gravity $\log(g)$, and the angular diameter $\Theta = 2R_\star/d$, which is a function of stellar radius $R_\star$ and distance $d$ and, hence, a distance scaling factor. Interstellar reddening is accounted for by applying a widely used extinction law\cite{1999PASP..111...63F}, which introduces the colour excess $E(B-V)$ as another free parameter (The extinction parameter $R_V$ is kept fixed at $3.1$, the value of the diffuse interstellar medium.). The best-fitting model has a reduced $\chi^2$ of $1.2$ and is shown in Fig.~\ref{fig:photometry}. The respective parameters are listed in Table~\ref{table:photometry} and go well with a mid B-type subgiant with zero interstellar reddening. There is no flux excess in the ultraviolet or in the infrared, that is, a luminous companion can be ruled out.
\subsection{Light curve}\label{subsection:light_curve}
\begin{figure}
\centering
\includegraphics[width=0.49\textwidth]{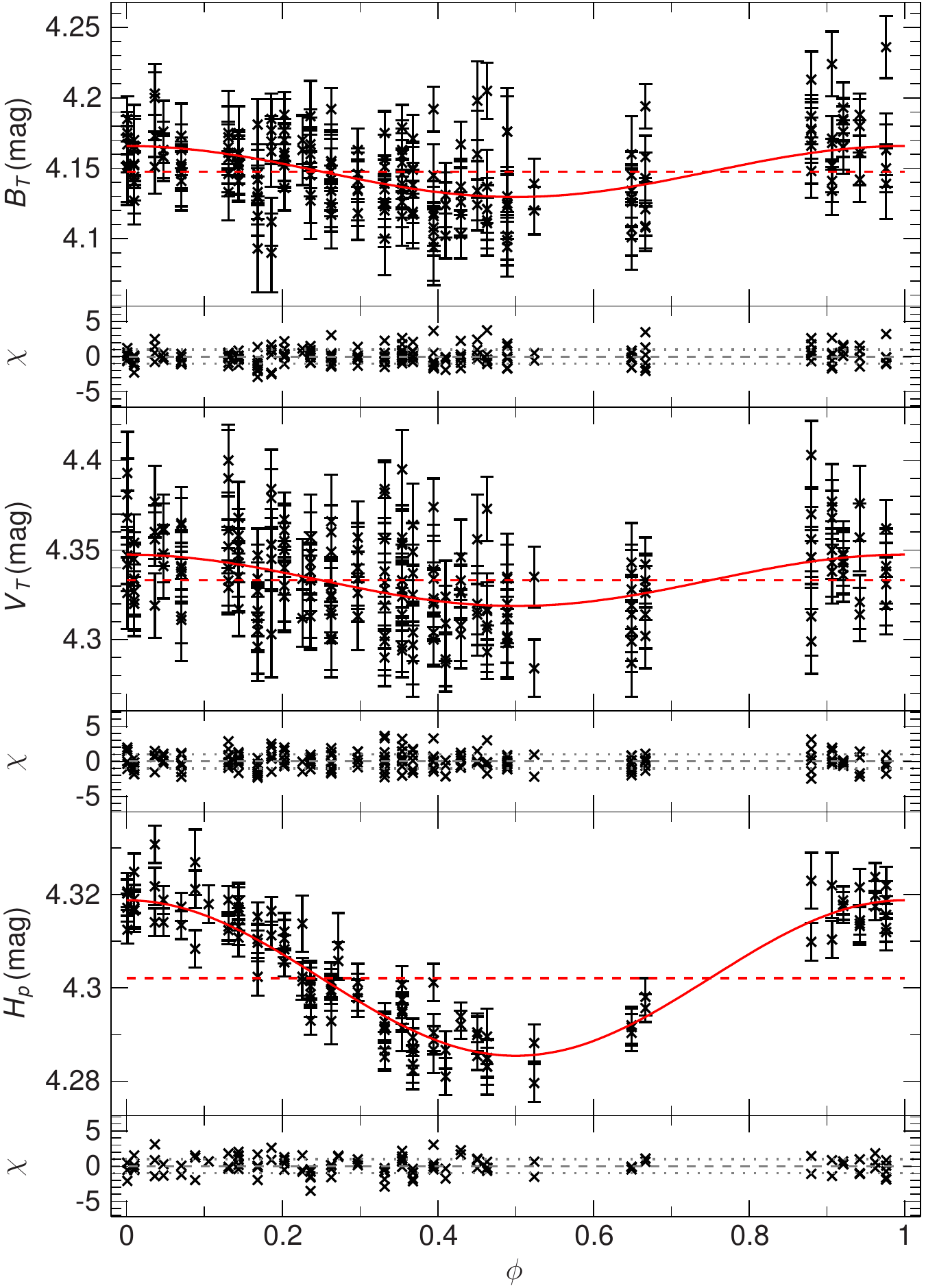}
\caption{\small Phased light curves of Tycho and {\sc Hipparcos} epoch photometry: the measurements are represented by black crosses with error bars (68\% confidence limits) while the best-fitting model (see Table~\ref{table:oscillation_params}) is indicated by the red solid curve. The red dashed line marks the derived mean magnitude. Residuals $\chi$ are shown as well.}
\label{fig:phased_lightcurves}
\end{figure}

\begin{figure}
\centering
\includegraphics[width=0.49\textwidth]{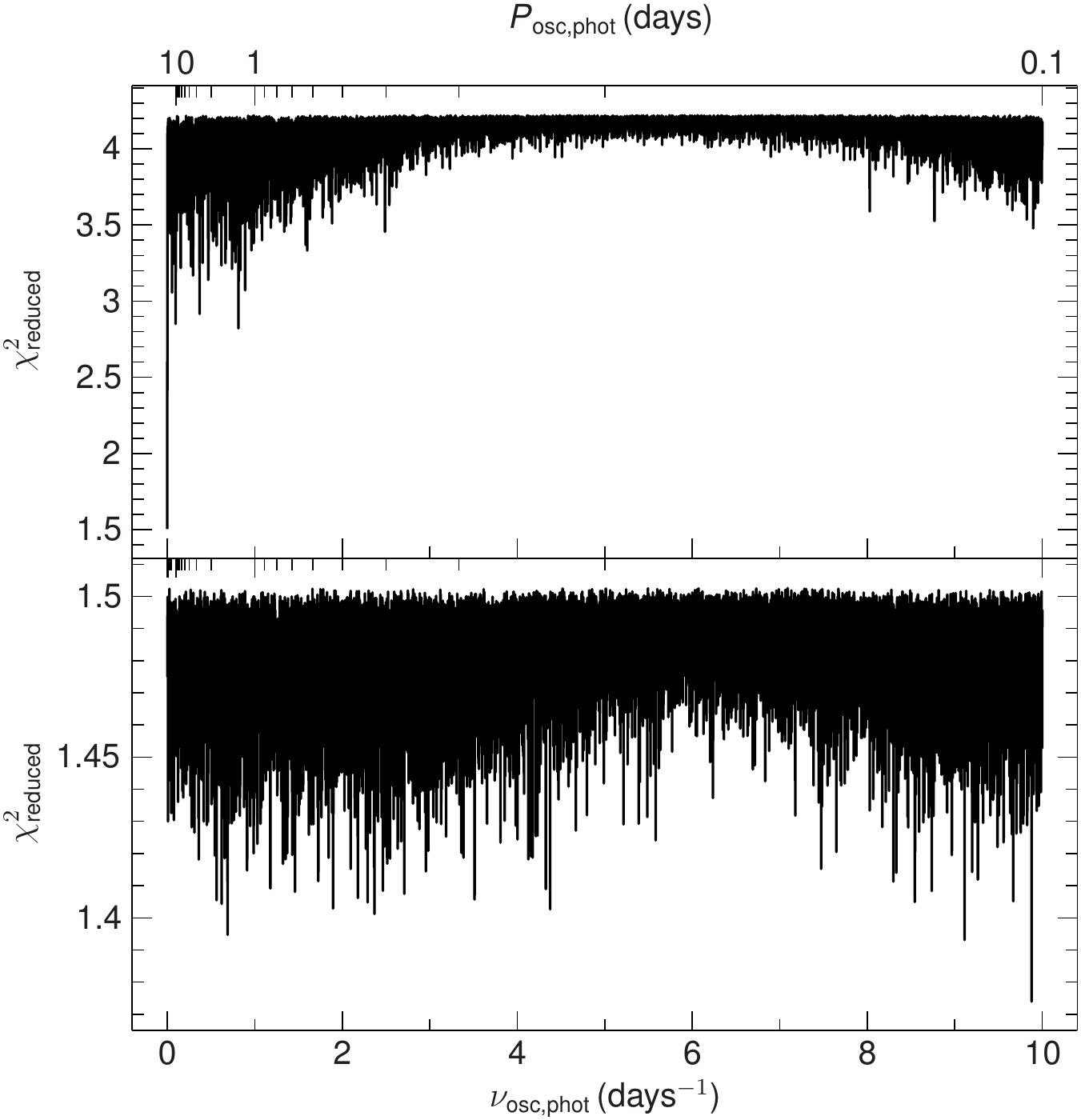}
\caption{\small The $\chi^2$ landscapes (``periodograms''), which result from fitting the Tycho and {\sc Hipparcos} epoch photometry data with the model given in Eq.~(\ref{eq:cosine_fit}) (\textit{top}) and a model with an additional oscillation term (\textit{bottom}), as functions of the original (\textit{top}) and additional (\textit{bottom}) oscillation frequency $\nu_{\textnormal{osc}}$, which are sampled in steps of $0.05/1196$\,d$^{-1}$ to ensure that phase shifts are always less than $0.05$.}
\label{fig:periodogram}
\end{figure}

Owing to distortions in the spectral line profiles, $\gamma$\,Columbae was classified as a candidate slowly pulsating B (SPB) star\cite{2006A&A...452..945T}. Because this class of pulsating stars is also known to show photometric variability\cite{1991A&A...246..453W}, we studied its Tycho and {\sc Hipparcos} epoch photometry data \cite{1997ESASP1200.....P} to see whether its light curve is consistent with an SPB nature or not\footnote{The recently observed {\it TESS}\cite{2015JATIS...1a4003R} light curve shows a complex, non-sinusoidal pattern with amplitudes $\lesssim 1$\,mmag, which calls for a more sophisticated asteroseismic modelling that is beyond the scope of this discovery paper.}. With $117$ photometric measurements in the {\sc Hipparcos} $H_p$-band (24 out of 141 measurements are flagged and thus not considered here) and $201$ data points in each of the two Tycho bands ($B_T$, $V_T$), all of which are spread over $1196$\,days in about 40 blocks, the sampling is quite sparse. Despite this, inspection by eye already indicates that the star exhibits a sinusoidal oscillation with a period of several years. To model this modulation, an analytic curve of the simplest form
\begin{equation}
\textnormal{mag}_j(t) = \overline{\textnormal{mag}}_j + A_{j} \cos\left(2\pi\left[(t-T_{\textnormal{ref}}) \nu_\textnormal{osc}+\phi_{\textnormal{ref}}\right] \right)
\label{eq:cosine_fit}
\end{equation}
with a time-dependent magnitude $\textnormal{mag}_j(t)$, a mean magnitude $\overline{\textnormal{mag}}_j$, an oscillation semiamplitude $A_{j}$, and an oscillation frequency $\nu_\textnormal{osc}$ is chosen. The parameter $\phi_{\textnormal{ref}}$ is the phase at the fixed reference epoch $T_{\textnormal{ref}}$. The index $j \in \{V_T, B_T, H_p\}$ refers to the three available passbands, which are simultaneously fitted using $\chi^2$ minimization tools provided by the Interactive Spectral Interpretation System\cite{2000ASPC..216..591H}. The resulting best-fitting model parameters are listed in Table~\ref{table:oscillation_params} and the respective phased light curves are shown in Fig.~\ref{fig:phased_lightcurves}. The reduced $\chi^2$ at the best fit is close to $1.5$, which indicates that the data points are not perfectly matched by our simplistic model. Adding additional frequencies, however, does not improve the quality of the fit. This is demonstrated in Fig.~\ref{fig:periodogram}, which shows the effect of adding a second oscillation term in Eq.~(\ref{eq:cosine_fit}). The reduced $\chi^2$ is only slightly lowered and still far off its desired value of $1$. Moreover, the shape of the resulting $\chi^2$ landscape does not hint at the presence of a second modulation because there is no distinct drop at a certain frequency, but rather resembles that of fitting a constant signal with noise. The slight mismatch indicated by the reduced $\chi^2$ may thus be caused by outliers and/or underestimated uncertainties in the measurements.

The derived semiamplitudes (15--20\,mmag) are in the typical range of SPB stars while the observed period ($1,520\pm140$\,d) seems to be incompatible with those objects that normally show oscillation periods of the order of days, not years\cite{2007CoAst.150..167D}. A possible explanation invoking an alias effect (caused, e.g., by regularities in the very low sampling rate) is ruled out here because the data cannot be fitted reasonably with periods of the order of days. Instead, this apparent contradiction can be easily resolved by noting that observed and intrinsic frequencies are different when the oscillating star is also rotating. For simplicity, the following three assumptions are usually made in order to derive the relation between the two frequencies: i) The star rotates uniformly with rotation frequency $\nu_\textnormal{rot}$. ii) The axes of rotation and pulsation are aligned. iii) The rotation frequency is small compared to the intrinsic oscillation frequency in the non-rotating case (labelled by the superscript ``$(0)$''): $\nu_\textnormal{rot}/\nu_\textnormal{osc}^{(0)} < 0.5$\cite{1997A&AS..121..343S}. The last assumption allows to neglect effects of the centrifugal force, which are proportional to $(\nu_\textnormal{rot}/\nu_\textnormal{osc}^{(0)})^2$, and to focus only on those of the Coriolis force, which are proportional to $\nu_\textnormal{rot}/\nu_\textnormal{osc}^{(0)}$. Under those three conditions, it can be shown\cite{1951ApJ...114..373L} that the rotationally induced change in the intrinsic frequency depends on the azimuthal order $m$ of the oscillation mode and on a structure constant $C$:%
\begin{equation}
\nu_\textnormal{osc}^\textnormal{obs} = \nu_\textnormal{osc}^{(0)} + m \nu_\textnormal{rot} C - m \nu_\textnormal{rot} = \nu_\textnormal{osc}^{(0)} + m \nu_\textnormal{rot} (C - 1) \,.
\label{eq:frequency_observed}
\end{equation}
The quantity $C$ can be approximately written as a function of the degree $l$ of the oscillation mode\cite{1997A&AS..121..343S}: $C=[2l(l+1)]^{-1}$. The first correction term in Eq.~(\ref{eq:frequency_observed}) is due to the Coriolis force while the second one is just a consequence of the alignment of the axes of rotation and pulsation: prograde ($m<0$) waves, i.e., waves traveling in direction of rotation, appear to move faster while retrograde ($m>0$) waves are seemingly slower. Because $C \le 1/4$ and thus $C-1 \le 0$, only certain combinations involving retrograde waves can lead to very low observed oscillation frequencies in Eq.~(\ref{eq:frequency_observed}). The fact that such a long-period oscillation mode is indeed observed for the program star is thus not only strong support for the axis-alignment assumption but also constrains the properties of this mode, because it has to have $m>0$. We conclude that the light curve is consistent with an SPB nature although the apparent oscillation frequency is probably the smallest ever observed for such an object.
\section{Line-profile analysis}\label{sect:line_profile}
\begin{figure*}
\centering
\includegraphics[angle=90,width=1\textwidth]{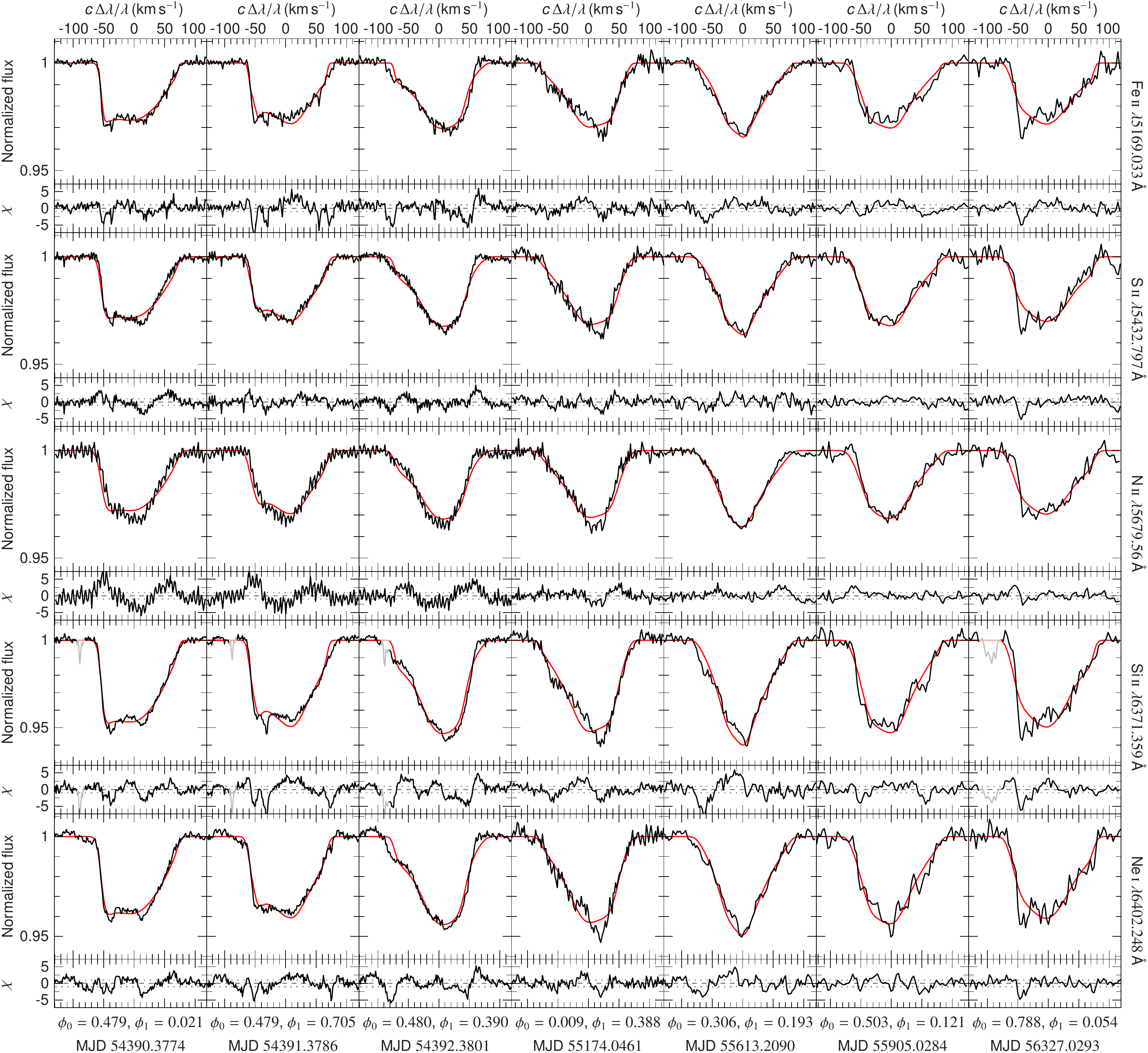}
\caption{\small Modelling of the pulsationally driven line-profile distortions for seven epochs (\textit{rows}) and five exemplary lines (\textit{columns}): the observations are indicated by a black line, the best-fitting model (see Sect.~\ref{sect:line_profile} for details) by a red one, and the quality of the fit by the residuals $\chi$. Regions excluded from fitting are shown in light colours. Phases $\phi$ with respect to the two observed oscillation frequencies are listed on the right-hand side.}
\label{fig:line_profile_variations}
\end{figure*}

In addition to the photometric variability, the star also exhibits strong temporal changes in the spectral line profiles of the order of days (see Fig.~\ref{fig:line_profile_variations}). Such a behaviour is expected for SPB stars because their non-radial oscillation modes can lead to surface velocity fields that significantly alter the underlying broadening profile caused by the Doppler effect induced by stellar rotation\cite{1997A&AS..121..343S}. In principle, two alternative explanations for the line profile distortions, namely moving spots on the surface and a double-lined spectroscopic binary system (SB2), exist, but are highly unlikely in this case because their spectral signature does not agree with what is observed here. Stellar spots manifest themselves as bumps that move across the spectral line profiles, with different behavior for lines of different chemical species\cite{2001A&A...380..177B,2004A&A...413..273B}. The distortions shown in Fig.~\ref{fig:line_profile_variations}, however, are almost identical for all elements and show vertical edges that are not bump-like at all. The blending of lines in an SB2 system is also not able to produce sharp edges that temporarily appear and then disappear again. Moreover, the measured parallaxes by {\sc Hipparcos} and {\it Gaia} (see Table~\ref{table:astrometry}) are inconsistent with the spectrophotometric distance inferred for a binary system composed of two stars with atmospheric parameters as indicated by our spectroscopic analysis (see Sect.~\ref{sect:spectroscopy}). Last but not least, both alternatives cannot provide a straightforward explanation for the simultaneous presence of a short-period line-profile variation and a long-period photometric variation. 

Stellar pulsations, in contrast, do: Our spectroscopic analysis (see Sect.~\ref{sect:spectroscopy}) places the star in the vicinity of the instability domain of SPB stars\cite{2016MNRAS.455L..67M}. Those objects are known to excite multiple modes. Therefore, the two observed frequencies are probably just caused by two distinct SPB-like pulsation modes with different properties. The long-period one observed in the light curve may be the result of a chance interplay between oscillation and rotation frequency as described by Eq.~(\ref{eq:frequency_observed}). The short-period one observed in the line profiles may not be detected in the light curve due to its low photometric amplitude, which, for instance, can be caused by cancellations effects in modes with high degree $l$. In order to test this hypothesis, we analysed the available high-quality spectra, which were taken at seven distinct epochs (see Table~\ref{table:available_spectra}), basically following the procedure that we already have employed to analyse the SPB star 18\,Pegasi\cite{2016A&A...591L...6I}. In contrast to the moment method\cite{2003A&A...398..687B} or the Fourier parameter fit method\cite{2006A&A...455..227Z}, this strategy is applicable to those cases where only a few observations are available. The underlying idea is to directly model the line profile variations in the spectra. To this end, we compute synthetic spectral line profiles based on the velocity field of a uniformly rotating, adiabatically pulsating star whose pulsation and rotation axes are aligned. For a single oscillation mode, the mathematical formulation\cite{1997A&AS..121..343S} of that field is a function of the degree $l$ and azimuthal order $m$ of the respective mode, the vertical amplitude $a_{\textnormal{sph}}$, the ratio of the horizontal and vertical amplitude $k^{(0)}$ (superscripts $^{(0)}$ refer to quantities in the non-rotating case), the ratio of the rotation frequency $\nu_\textnormal{rot}$ and the oscillation frequency $\nu_\textnormal{osc}^{(0)}$, the inclination of the pulsational/rotational axis $i$, the projected rotational velocity $v\sin(i)$, and the observed oscillation phase $\phi_\textnormal{osc}^\textnormal{obs}$. Assuming a linear pulsation theory, extension to multiple modes can simply be achieved by adding the velocity fields of all modes together. Numerical integration of the resulting velocity field over the projected stellar disk (using a standard linear limb-darkening law) allows to synthesize the Doppler broadening profiles. We note that this approach is only valid in the slow-rotation case where $\nu_\textnormal{rot}/\nu_\textnormal{osc}^{(0)} < 0.5$ because effects of the Coriolis force are included in the mathematical description while those of the Centrifugal force are not (see Sect.~\ref{subsection:light_curve}). A further limitation is given by the fact that the formulation is purely dynamical, which means that it does not account for the effects of pulsationally driven changes in the atmospheric parameters on the local surface brightness. Because their intrinsic line profiles are almost perfectly Gaussian, we focus on isolated, narrow metal lines when fitting the observations with our model, which consists of Gaussian absorption lines (parameterized by area, width, and fixed central wavelength) that are convolved with the pulsational broadening profile as well as with the respective instrumental profile. For simplicity (and consistency with the model for the velocity field), the intrinsic lines shapes (i.e., area and width of each Gaussian) are kept fixed with time, which again means that atmospheric changes caused by pulsations are neglected. For each observation date, a radial-velocity parameter $v_\textnormal{rad}$ is introduced to allow for possible Doppler shifts in the central wavelengths. Individual oscillation phases are not fitted independently but coupled via a parameter that represents the observed oscillation frequency (as given in Eq.~(\ref{eq:frequency_observed})). Using a model with two oscillation modes, five metal lines from five different chemical species are simultaneously fitted. Unfortunately, numerical optimization turns out to be a difficult task owing to the large number of free parameters and the complicated $\chi^2$ landscape. Despite huge efforts, there is no guarantee that the global best fit is indeed found. To simplify the optimization procedure, the size of the multi-parameter space is reduced by demanding that stellar radii, masses and surface gravities shall be within the intervals $4\,R_\odot \lesssim R_\star \lesssim 12\,R_\odot$, $2\,M_\odot \lesssim M \lesssim 10\,M_\odot$, and $2.7 \lesssim \log(g) \lesssim 3.9$, respectively. Moreover, the allowed range for the observed frequency of the first oscillation mode is restricted to values around the result of the light-curve analysis (see Table~\ref{table:oscillation_params}). Finally, all physically possible combinations between the degrees $l$ and azimuthal orders $m$ of the two modes up to $l=6$ are explored separately rather than fitting these four parameters directly. 

Table~\ref{table:pulsation_modeling} lists the most relevant free parameters and derived quantities for the $30$ best-fitting combinations. Given the various simplifications in the model and the superb quality of the available spectra, a perfect match with a reduced $\chi^2$ close to unity cannot be expected. Figure~\ref{fig:line_profile_variations} shows that the resulting best-fitting model, which consists of modes with $l=3$, $m=+3$ and $l=4$, $m=-1$ (see Table~\ref{table:pulsation_modeling}), is indeed able to reproduce the data. Because the properties of both fitted oscillation modes are in line with those of SPB stars, we conclude that $\gamma$\,Columbae belongs to this group of stars. Two points are important to stress here: Firstly, the radial velocity of the star does not seem to vary significantly. Secondly, the line-profile modelling allows us to estimate the stellar radius $R_\star$, mass $M$, and surface gravity $g$ (see notes of Table~\ref{table:pulsation_modeling} for details). For instance, the best-fitting model predicts $R_\star=7.1\,R_\odot$, $M=4.0\,M_\odot$, and $\log(g)=3.34$.
\section{Spectroscopic analysis}\label{sect:spectroscopy}
The spectroscopic analysis closely follows our standard procedure\cite{2014A&A...565A..63I}. In short, all available spectra (see Table~\ref{table:available_spectra}) are fitted simultaneously over their entire spectral range to deduce the atmospheric parameters and the abundances of the individual chemical species. Model spectra are based on atmospheric structures in local thermodynamic equilibrium (LTE) computed with the {\sc Atlas12} code\cite{1996ASPC..108..160K}. Departures from LTE are accounted for by making use of updated versions of {\sc Detail} and {\sc Surface}\cite{1981PhDT.......113G,detailsurface2}. The {\sc Detail} code computes population numbers in non-LTE while the {\sc Surface} code uses these numbers and more detailed line-broadening data to compute the final synthetic spectrum. This so-called hybrid LTE/non-LTE approach has been recently improved\cite{2018A&A...615L...5I} by allowing for non-LTE effects on the atmospheric structure as well as by the implementation of the occupation probability formalism\cite{1994A&A...282..151H} for hydrogen and new Stark broadening tables for hydrogen\cite{2009ApJ...696.1755T} and neutral helium\cite{1997ApJS..108..559B}. Pulsationally driven line-profile variations are modelled in the same way as described in Sect.~\ref{sect:line_profile}, with the corresponding parameters (except the projected rotational velocity and radial velocities) being fixed at the values of the respective best-fitting model from Table~\ref{table:pulsation_modeling}. Figure \ref{fig:spectra_1} shows the comparison between the observed ESPaDOnS spectrum and the best-fitting model spectrum, the atmospheric parameters and abundances of which are listed in Table~\ref{table:spectroscopy}. The atmospheric parameters ($T_{\mathrm{eff}}=15,570$\,K, $\log(g)=3.328$) agree almost perfectly with those from photometry and line-profile modelling and resemble those of mid B-type subgiants. According to Fig.~\ref{fig:abundance_pattern}, the baseline metallicty of the star is characteristic of normal B-type stars in the solar neighbourhood\cite{2012A&A...539A.143N}, while abundances of He, C, N, and O show signatures for hydrogen burning via the CNO bi-cycle that are far too strong to be explained by effects of normal stellar evolution, e.g., via rotational mixing. For instance, solar-metallicity, single-star models with masses between 4 to 7\,$M_\odot$ that start their evolution with a surface rotation near the critical limit are expected to show enhancements in the N/H surface abundance ratio by factors between 3 and 5$^{\textnormal{Fig.~8 in}}$\cite{2013A&A...553A..24G}. These values can be seen as upper limits which may only be reached by initially very fast rotating stars. Because the N/H surface ratio in $\gamma$\,Columbae is enhanced by a factor of $11.4^{+7.6}_{-4.8}$ (99\% confidence interval) with respect to the respective solar value, we conclude that this strong enrichment cannot be attributed to rotational mixing alone.
\section{Astrometric analysis}\label{sect:astrometry}
Besides its peculiar abundance pattern, additional evidence for deviations from standard single-star evolution of $\gamma$\,Columbae arise when looking at its parallax, which was measured by {\sc Hipparcos} as well as by {\it Gaia}. Because both measurements agree exceptionally well (see Table~\ref{table:astrometry}) and because the most meaningful {\it Gaia} quality flag for bright blue stars, namely the ``renormalized unit weight error'' (RUWE)\cite{RUWE}, indicates a well-behaved astrometric solution ($\mathrm{RUWE}=1.04$), we consider it to be highly trustworthy. We use the {\sc Hipparcos} parallax $\varpi$ --~which is still more precise than the currently available one from {\it Gaia}'s second data release~-- to estimate the stellar radius and mass. To this end, the definitions for the angular diameter ($\Theta=2R_\star/d = 2R_\star \varpi$, $d$ is the distance) and for the surface gravity ($g=GM/R_\star^2$, $G$ is the gravitational constant) have to be combined. Using $\Theta$ from Table~\ref{table:photometry} as well as $T_{\mathrm{eff}}$ and $\log(g)$ from Table~\ref{table:spectroscopy}, we obtain $M=5.0\pm1.7\,M_\odot$, $R_\star=8.0\pm1.0\,R_\odot$, and $\log(L/L_\odot)=3.53\pm0.12$ (99\% confidence intervals), which is in good agreement with the best-fitting model for the line-profile variations, $M=4.0\,M_\odot$ and $R_\star = 7.1\,R_\odot$ (see Table~\ref{table:pulsation_modeling}), but, in terms of mass, in tension with theoretical predictions for single-star evolution, which according to Fig.~\ref{fig:evolution_tracks}, suggest a higher mass of about $7\,M_\odot$.

In order to investigate whether the star shows peculiar kinematic behaviour, e.g., caused by a strong dynamical encounter, past trajectories of $\gamma$\,Columbae are computed in a Galactic mass model\cite{2013A&A...549A.137I}. Figure~\ref{fig:kinematics} shows nine orbits whose initial conditions sample the typical range given by the uncertainties of the input parameters. The shape of the trajectories is characteristic of thin-disk stars in the solar neighbourhood and, thus, completely inconspicuous.

{\textbf{Data availability}
The astrometric, photometric, and spectroscopic data used in this work are all publicly available.
}
\begingroup
\renewcommand{\section}[2]{}
\setlength{\bibsep}{0pt}

\endgroup
\clearpage
\onecolumn
\section*{Extended data}
\begin{figure*}
\centering
\includegraphics[height=1\textwidth, angle=-90]{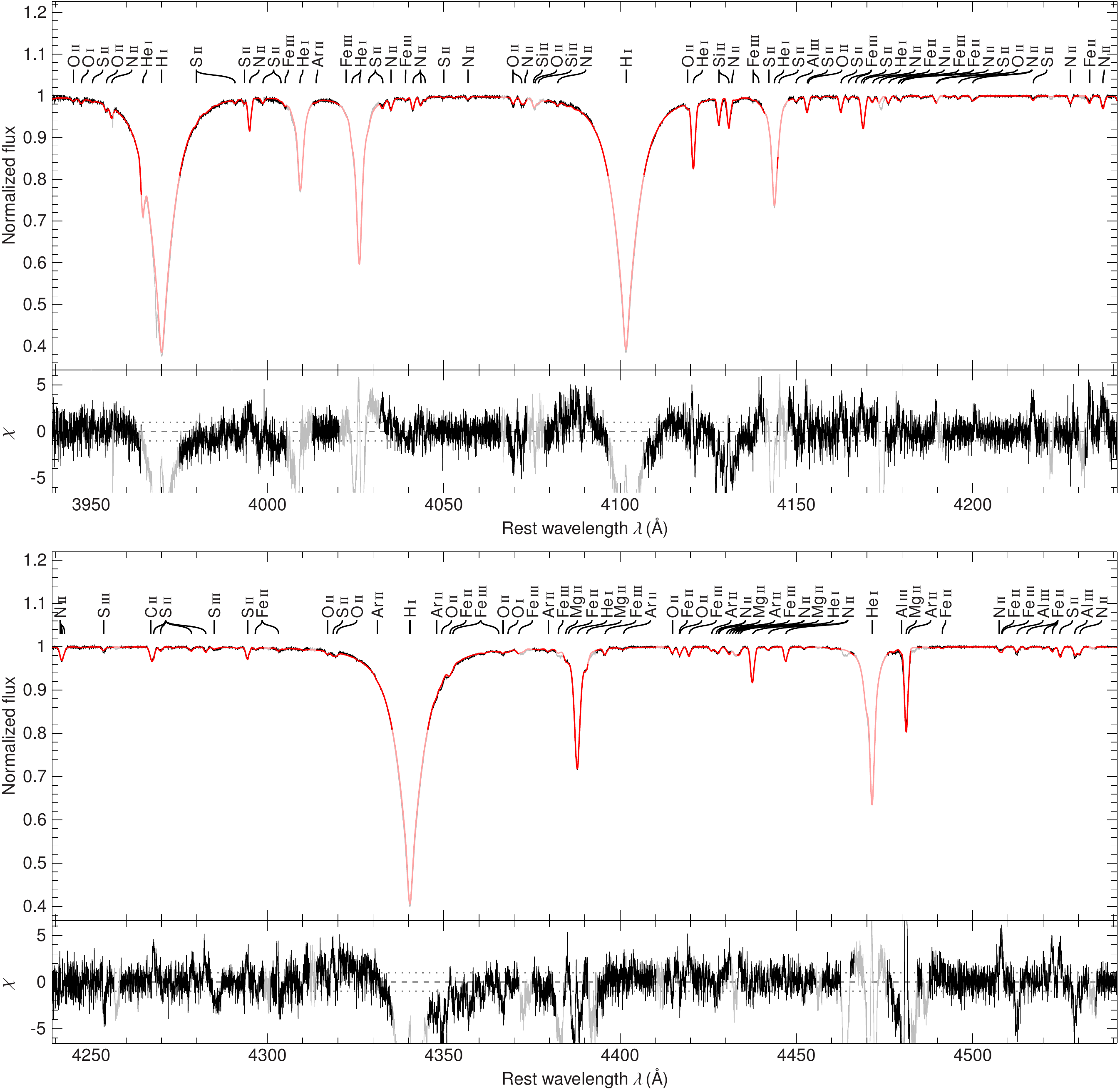}
\caption{\small Comparison of best-fitting model spectrum (red line) with re-normalized observation (black line). Light colours mark regions that have been excluded from fitting (e.g., due to data reduction artefacts or the presence of features that are not properly included in our models). For the sake of clarity, only the strongest out of all lines that have been used in the analysis are labelled. Residuals $\chi$ are shown as well. Telluric correction is performed via interpolation within a pre-calculated grid of transmission spectra\cite{2014A&A...568A...9M}. Although the atmospheric conditions used in that spectral library are tailored to Cerro Paranal, the two free parameters airmass and precipitable water vapour content are enough to ensure a decent representation of many telluric features for different observing sites and weather conditions. Telluric features that are not properly reproduced by this approach have been excluded as well.}
\label{fig:spectra_1}
\end{figure*}
\addtocounter{figure}{-1}
\begin{figure*}
\centering
\includegraphics[height=1\textwidth, angle=-90]{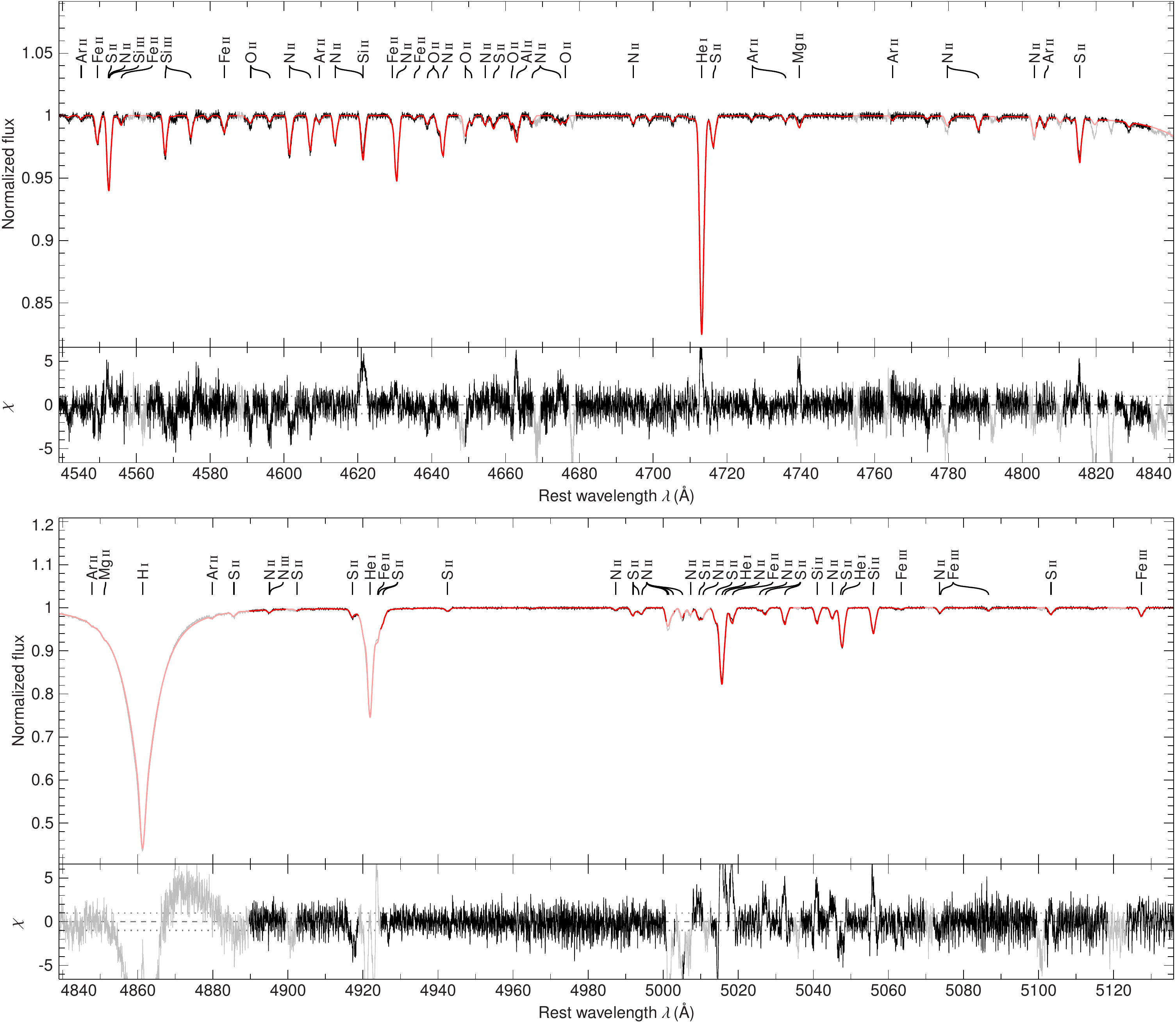}
\caption{\small continued.}
\end{figure*}
\addtocounter{figure}{-1}
\begin{figure*}
\centering
\includegraphics[height=1\textwidth, angle=-90]{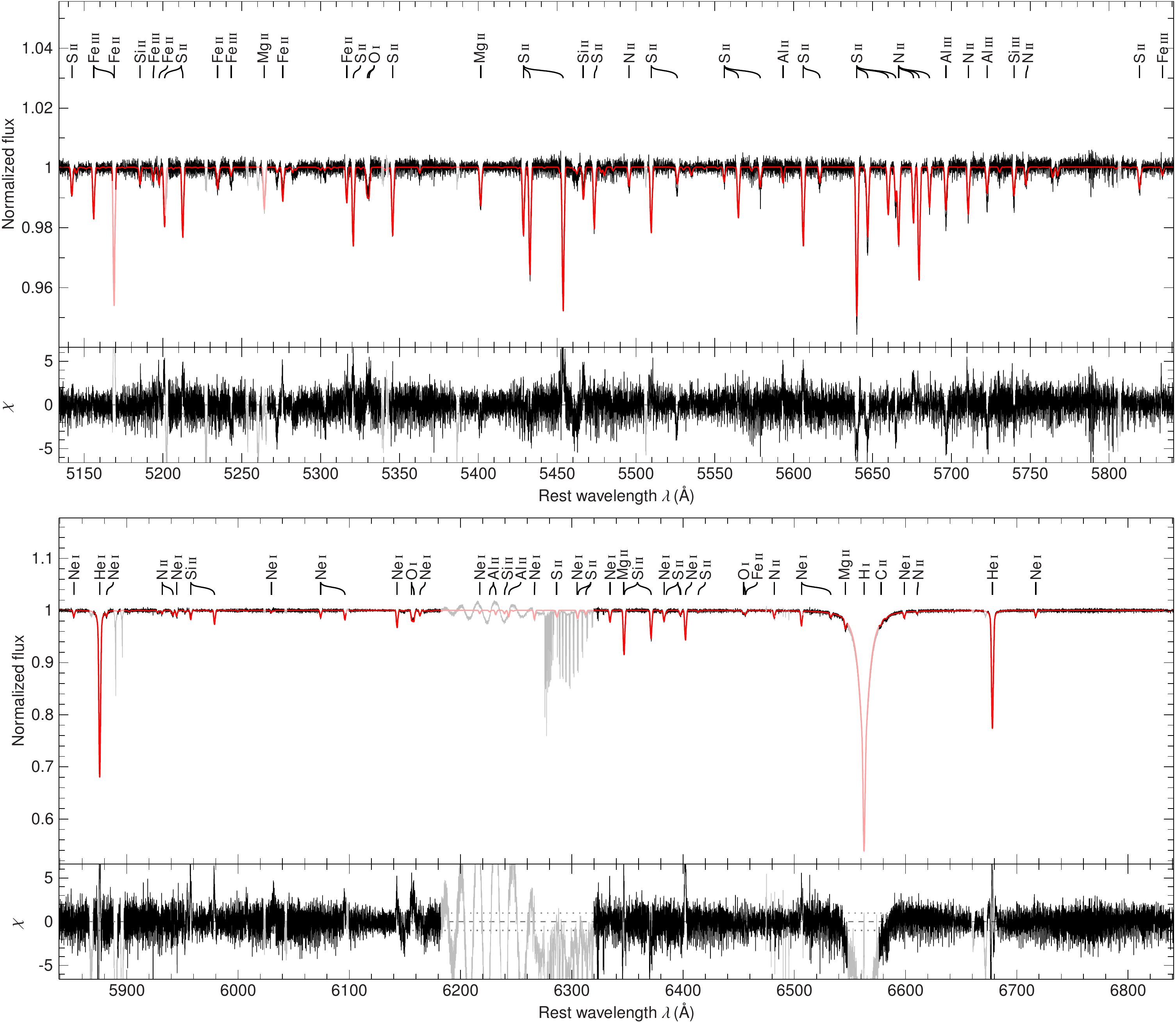}
\caption{\small continued.}
\end{figure*}
\addtocounter{figure}{-1}
\begin{figure*}
\centering
\includegraphics[height=1\textwidth, angle=-90]{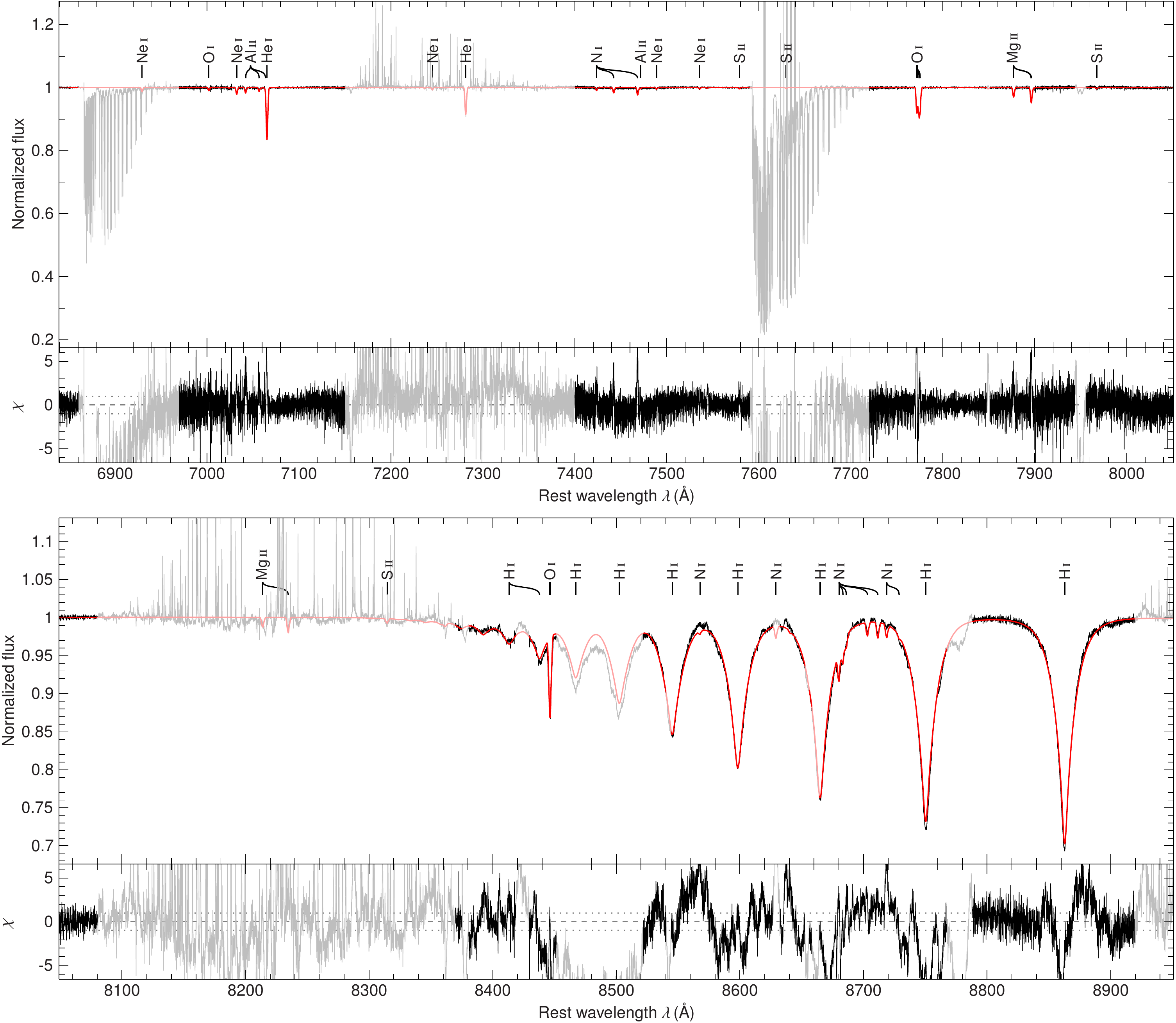}
\caption{\small continued.}
\end{figure*}
\begin{figure}
\centering
\includegraphics[width=0.49\textwidth]{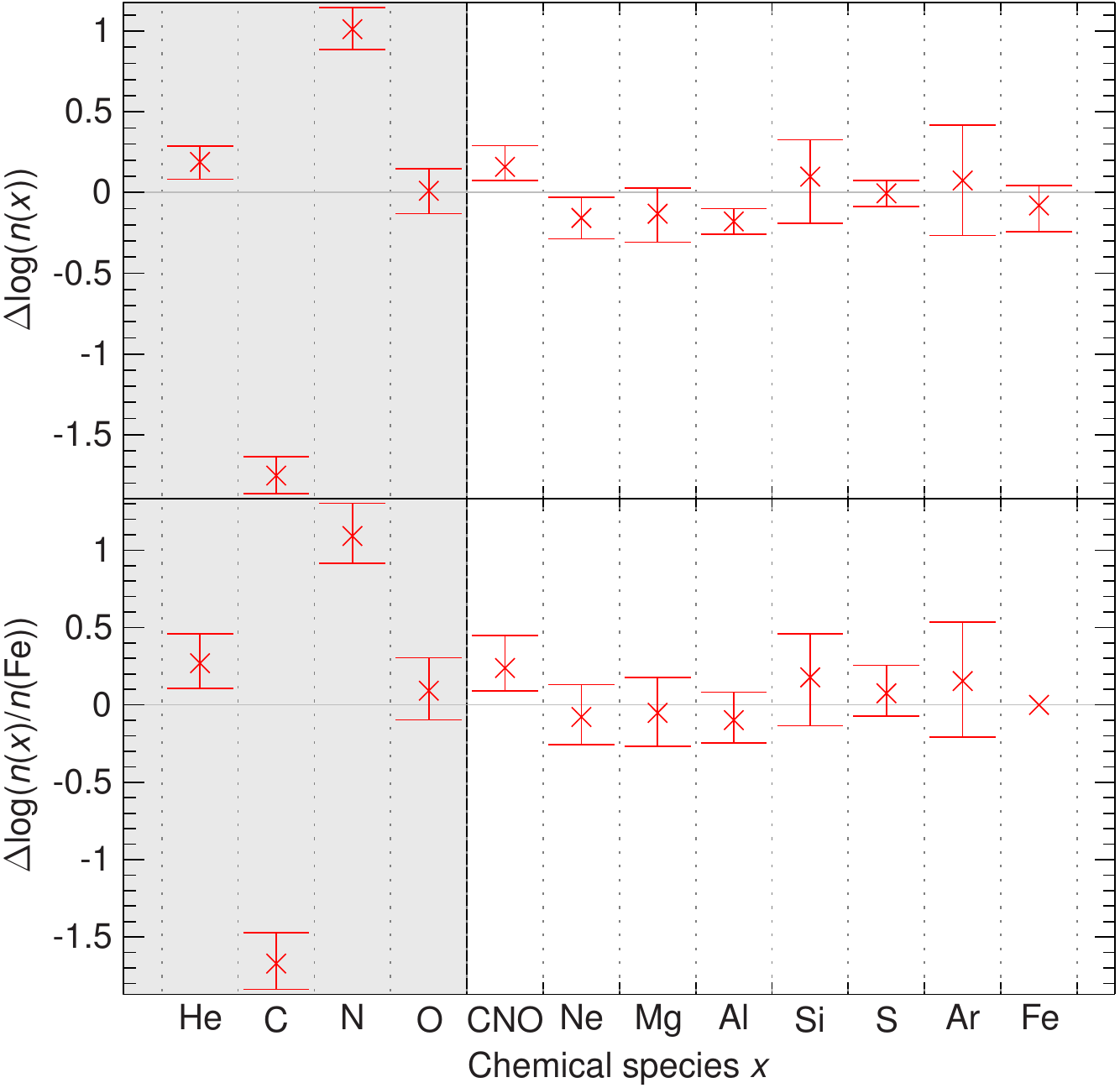}
\caption{\small Differential abundance pattern (\textit{top}) and element-over-iron abundance ratios (\textit{bottom}) of $\gamma$\,Columbae with respect to the cosmic abundance standard\cite{2012A&A...539A.143N}. Solar abundances\cite{2009ARA&A..47..481A} are used for Al, S, and Ar because cosmic abundances are not available for these elements. The error bars are the square roots of the quadratic sums of all given individual uncertainties and cover 99\% confidence intervals. The gray-shaded area marks elements whose abundances show signatures of hydrogen burning via the CNO bi-cycle. To get rid of this effect, the invariant sum of the bi-cycle-catalysts C, N, and O (``CNO'') is plotted, too.}
\label{fig:abundance_pattern}
\end{figure}
\begin{figure}
\centering
\includegraphics[width=0.49\textwidth]{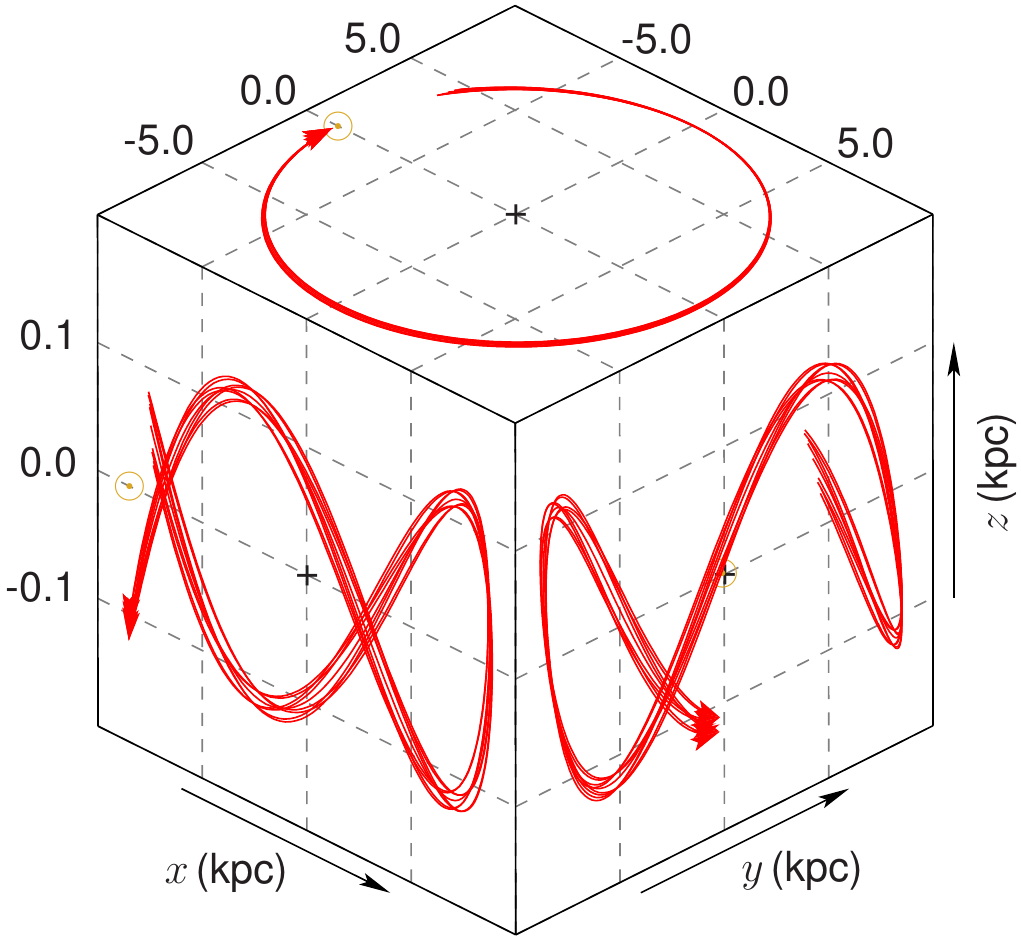}
\caption{\small Three-dimensional orbits in a Galactic Cartesian coordinate system in which the $z$-axis points to the Galactic north pole. The nine trajectories (red lines; arrows indicate the star's current position) are computed for the past $200$\,Myr and sample the mean kinematic input data as well as uncertainties in the parallax, proper motions, and radial velocity. The positions of the Sun and the Galactic center are marked by a yellow circled dot ($\odot$) and a black plus sign ($+$), respectively.}
\label{fig:kinematics}
\end{figure}
\clearpage
\begin{table}
\small
\caption{Stellar parameters derived from photometry.}
\label{table:photometry}
\begin{center}
\begin{tabular}{lr}
\hline\hline
Parameter & Value \\
\hline
Angular diameter $\Theta$ & $(1.351\pm 0.025)\times10^{-9}$\,rad \\
Colour excess $E(B-V)$ & $\le0.017$\,mag \\
Effective temperature $T_{\mathrm{eff}}$ & $15,500^{+600}_{-400}$\,K \\
Surface gravity $\log (g\,\mathrm{(cm\,s^{-2})})$ & $3.6\pm0.5$ \\
\hline
\end{tabular}
\end{center}
{\bf Notes:} The given uncertainties are single-parameter 99\% confidence intervals based on $\chi^2$ statistics around the best fit with $\chi^2_\textnormal{reduced} \approx 1.2$.
\end{table}
\begin{table}
\small
\caption{Observed oscillation parameters derived from Tycho and {\sc Hipparcos} epoch photometry data.}
\label{table:oscillation_params}
\begin{center}
\begin{tabular}{lr}
\hline\hline
Parameter & Value \\
\hline
Frequency $\nu_{\textnormal{osc}}^{\textnormal{obs}}$ & $0.00066\pm0.00006$\,d$^{-1}$ \\
Period $P_{\textnormal{osc}}^{\textnormal{obs}}$ & $1520\pm140$\,d \\
Reference epoch $T_{\textnormal{ref}}$ (fixed) & $47,857.0$\,MJD \\
Phase $\phi_{\textnormal{ref}}$ at epoch $T_{\textnormal{ref}}$ & $0.88\pm0.04$ \\
$B_T$ mean magnitude & $4.148\pm0.004$\,mag \\
$B_T$ semiamplitude & $18\pm5$\,mmag \\
$V_T$ mean magnitude & $4.333\pm0.004$\,mag \\
$V_T$ semiamplitude & $14\pm5$\,mmag \\
$H_p$ mean magnitude & $4.3021\pm0.0012$\,mag \\
$H_p$ semiamplitude & $16.7\pm1.3$\,mmag \\
\hline
\end{tabular}
\end{center}
{\bf Notes:} The given uncertainties are single-parameter 99\% confidence intervals based on the $\chi^2$ statistics around the best fit with $\chi^2_\textnormal{reduced} \approx 1.5$.
\end{table}
\begin{table}
\small
\caption{Astrometric data.}
\label{table:astrometry}
\begin{center}
\begin{tabular}{lrr}
\hline\hline
Parameter & \multicolumn{2}{c}{Value} \\
\cline{2-3}
& {\sc Hipparcos}{\cite{2007A&A...474..653V}} & {\it Gaia} DR2{\cite{2018A&A...616A...1G}} \\
\hline
Parallax $\varpi$ (mas) & $3.75\pm0.18$ & $3.73\pm0.32$ \\
Proper motion $\mu_\alpha \cos\delta$ (mas\,yr$^{-1}$) & $-3.24\pm0.16$ & $-3.63\pm0.51$ \\
Proper motion $\mu_\delta$ (mas\,yr$^{-1}$) & $10.21\pm0.16$ & $10.50\pm0.50$ \\
\hline
\end{tabular}
\end{center}
{\bf Notes:} The given uncertainties are assumed to be 68\% confidence intervals. 
\end{table}
\begin{table}
\small
\caption{Mid-exposure modified Julian date, spectral resolution, and median signal-to-noise ratio of the available spectra.}
\label{table:available_spectra}
\begin{center}
\begin{tabular}{rrlr}
\hline\hline
MJD & $R = \lambda/\Delta \lambda$ &  S/N & Spectrograph \\
\hline
$54\,390.3774$ & $107\,200$ & $770^{\textnormal{a}}$ & {\sc Uves}\cite{2000SPIE.4008..534D} \\
$54\,391.3786$ & $107\,200$ & $750^{\textnormal{a}}$ & {\sc Uves}\cite{2000SPIE.4008..534D} \\
$54\,392.3801$ & $107\,200$ & $730^{\textnormal{a}}$ & {\sc Uves}\cite{2000SPIE.4008..534D} \\
$55\,174.0461$ &  $80\,000$ & $400^{\textnormal{b}}$ & {\sc Harps}\cite{2003Msngr.114...20M} \\
$55\,613.2090$ &  $65\,000$ & $620^{\textnormal{c}}$ & ESPaDOnS\cite{2006ASPC..358..362D} \\
$55\,905.0284$ &  $48\,000$ & $300$                  & {\sc Feros}\cite{1999Msngr..95....8K} \\
$56\,327.0293$ &  $48\,000$ & $270$                  & {\sc Feros}\cite{1999Msngr..95....8K} \\
\hline
\end{tabular}
\end{center}
{\bf Notes:} ${}^{\textnormal{(a)}}$Coadded from 3 individual spectra;  ${}^{\textnormal{(b)}}$Coadded from 5 individual spectra;  ${}^{\textnormal{(c)}}$Coadded from 9 individual spectra.
\end{table}
\begin{sidewaystable*}
\small
\setlength{\tabcolsep}{0.069cm}
\renewcommand{\arraystretch}{1.25}
\caption{Results of the line-profile modelling for the 30 best-fitting combinations of the two oscillation modes.}
\label{table:pulsation_modeling}
\begin{center}
\footnotesize
\begin{tabular}{r|rrrrrr|rrrr|rrrrrrrrr|rr|rrrr|rrrrrr}
\hline\hline
& \multicolumn{19}{c|}{Most relevant parameters affecting} & \multicolumn{11}{c}{Derived quantities from} \\
& \multicolumn{6}{c|}{the first mode} & \multicolumn{4}{c|}{the second mode} & \multicolumn{9}{c|}{both modes} & \multicolumn{2}{c|}{the first mode} & \multicolumn{4}{c|}{the second mode} & \multicolumn{5}{c}{both modes} \\
\cline{2-7} \cline{8-11} \cline{12-20} \cline{21-22} \cline{23-26} \cline{27-31}
$\chi^2_\textnormal{red}$ & $l$ & $m$ & ${\langle v_\mathrm{v}^2 \rangle}{}^{1/2}$ & $k^{(0)}$ & $\nu_\textnormal{rot}/\nu_\textnormal{osc}^{(0)}$ & $\nu_\textnormal{osc}^\textnormal{obs}$ & $l$ & $m$ & ${\langle v_\mathrm{v}^2 \rangle}{}^{1/2}$ & $\nu_\textnormal{osc}^\textnormal{obs}$ & $i$ & $v\sin(i)$ & \multicolumn{7}{c|}{$v_\textnormal{rad}$ at the 7 distinct epochs} & $a_{\mathrm{sph}}$ & $\nu_\textnormal{osc}^{(0)}$ & $k^{(0)}$ & $\nu_\textnormal{rot}/\nu_\textnormal{osc}^{(0)}$ & $a_{\mathrm{sph}}$ & $\nu_\textnormal{osc}^{(0)}$  & $\eta$ & $\nu_\textnormal{rot}$ & $M$ & $R_\star$ & $\log(g)$ \\
\cline{14-20}
& & & ($\mathrm{km\,s^{-1}}$) & & & (d$^{-1}$) & & & ($\mathrm{km\,s^{-1}}$) & (d$^{-1}$) & ($^{\circ}$) & ($\mathrm{km\,s^{-1}}$) & \multicolumn{7}{c|}{($\mathrm{km\,s^{-1}}$)} & ($R_\odot$) & (d$^{-1}$) & & & ($R_\odot$) & (d$^{-1}$) & & (d$^{-1}$) & ($M_\odot$) & ($R_\odot$) & (cgs)\\
\hline
$3.42$ & $3$ & $+3$ & $ 0.92$ & $ 1.28$ & $0.348$ & $0.000671$ & $4$ & $-1$ & $ 6.80$ & $1.682$ & $40.4$ & $ 65.38$ & $ 21.1$ & $ 19.5$ & $ 21.4$ & $ 17.4$ & $ 19.1$ & $ 13.4$ & $ 21.8$ & $0.11$ & $0.807$ & $ 0.42$ & $0.199$ & $0.48$ & $1.409$ & $0.09$ & $0.281$ & $ 4.0$ & $ 7.1$ & $3.34$ \\ 
$3.53$ & $4$ & $+4$ & $ 0.36$ & $ 0.76$ & $0.256$ & $0.000668$ & $4$ & $+0$ & $ 2.62$ & $1.751$ & $30.6$ & $ 66.56$ & $ 20.1$ & $ 17.1$ & $ 18.9$ & $ 16.1$ & $ 17.6$ & $ 13.0$ & $ 19.1$ & $0.02$ & $1.823$ & $ 0.82$ & $0.267$ & $0.15$ & $1.751$ & $0.09$ & $0.467$ & $ 5.7$ & $ 5.5$ & $3.71$ \\ 
$3.57$ & $6$ & $+4$ & $ 0.10$ & $ 3.13$ & $0.253$ & $0.000638$ & $4$ & $+2$ & $ 9.69$ & $1.082$ & $78.9$ & $ 55.45$ & $ 17.4$ & $ 18.5$ & $ 20.1$ & $ 17.4$ & $ 17.0$ & $ 12.3$ & $ 18.9$ & $0.01$ & $0.663$ & $ 0.69$ & $0.119$ & $0.68$ & $1.409$ & $0.02$ & $0.168$ & $ 5.5$ & $ 6.7$ & $3.53$ \\ 
$3.61$ & $5$ & $+4$ & $ 0.15$ & $ 0.95$ & $0.254$ & $0.000667$ & $4$ & $-1$ & $10.13$ & $1.373$ & $49.6$ & $ 61.85$ & $ 19.5$ & $ 18.9$ & $ 19.1$ & $ 17.6$ & $ 20.2$ & $ 11.4$ & $ 20.7$ & $0.02$ & $0.810$ & $ 0.46$ & $0.176$ & $0.86$ & $1.172$ & $0.07$ & $0.206$ & $ 4.0$ & $ 7.8$ & $3.25$ \\ 
$3.69$ & $5$ & $+4$ & $ 0.39$ & $ 0.53$ & $0.254$ & $0.000639$ & $3$ & $+0$ & $ 3.40$ & $1.751$ & $20.8$ & $ 65.36$ & $ 23.8$ & $ 19.9$ & $ 16.3$ & $ 13.7$ & $ 17.6$ & $ 15.1$ & $ 23.1$ & $0.02$ & $2.080$ & $ 0.74$ & $0.302$ & $0.19$ & $1.751$ & $0.12$ & $0.529$ & $10.0$ & $ 6.9$ & $3.76$ \\ 
$3.74$ & $6$ & $+5$ & $ 0.12$ & $ 2.04$ & $0.202$ & $0.000657$ & $4$ & $+2$ & $ 9.97$ & $1.082$ & $79.1$ & $ 54.96$ & $ 17.6$ & $ 18.7$ & $ 19.3$ & $ 17.5$ & $ 17.9$ & $ 13.6$ & $ 20.4$ & $0.01$ & $0.816$ & $ 0.69$ & $0.118$ & $0.70$ & $1.404$ & $0.02$ & $0.165$ & $ 5.5$ & $ 6.7$ & $3.53$ \\ 
$3.75$ & $3$ & $+3$ & $ 2.57$ & $ 0.48$ & $0.348$ & $0.000638$ & $3$ & $+0$ & $ 4.80$ & $2.400$ & $15.9$ & $ 66.53$ & $ 21.1$ & $ 20.8$ & $ 14.2$ & $ 17.5$ & $ 17.2$ & $ 18.0$ & $ 26.0$ & $0.09$ & $2.737$ & $ 0.63$ & $0.397$ & $0.20$ & $2.400$ & $0.25$ & $0.952$ & $ 6.2$ & $ 5.0$ & $3.83$ \\ 
$3.75$ & $5$ & $+5$ & $ 1.95$ & $ 0.33$ & $0.203$ & $0.000672$ & $3$ & $+0$ & $ 4.85$ & $2.286$ & $22.8$ & $ 65.44$ & $ 22.6$ & $ 21.3$ & $ 13.0$ & $ 13.7$ & $ 17.1$ & $ 18.0$ & $ 20.5$ & $0.06$ & $2.955$ & $ 0.55$ & $0.263$ & $0.21$ & $2.286$ & $0.12$ & $0.601$ & $ 6.7$ & $ 5.5$ & $3.77$ \\ 
$3.77$ & $5$ & $+3$ & $ 0.18$ & $ 2.27$ & $0.339$ & $0.000676$ & $4$ & $-1$ & $ 8.99$ & $2.395$ & $45.4$ & $ 61.19$ & $ 20.2$ & $ 18.5$ & $ 18.8$ & $ 16.6$ & $ 20.9$ & $ 14.3$ & $ 19.3$ & $0.02$ & $0.951$ & $ 0.47$ & $0.155$ & $0.43$ & $2.081$ & $0.05$ & $0.322$ & $ 4.0$ & $ 5.3$ & $3.60$ \\ 
$3.80$ & $4$ & $+3$ & $ 0.43$ & $ 0.46$ & $0.342$ & $0.000639$ & $3$ & $+0$ & $ 3.46$ & $2.339$ & $13.4$ & $ 65.52$ & $ 22.6$ & $ 20.9$ & $ 13.0$ & $ 12.9$ & $ 15.5$ & $ 16.0$ & $ 24.2$ & $0.01$ & $2.918$ & $ 0.72$ & $0.426$ & $0.15$ & $2.339$ & $0.25$ & $0.997$ & $ 9.3$ & $ 5.6$ & $3.91$ \\ 
$3.87$ & $3$ & $+3$ & $ 1.28$ & $ 0.48$ & $0.348$ & $0.000638$ & $3$ & $-3$ & $ 8.45$ & $2.199$ & $17.5$ & $ 67.42$ & $ 21.3$ & $ 16.7$ & $ 20.9$ & $ 17.1$ & $ 17.4$ & $ 12.5$ & $ 17.0$ & $0.09$ & $1.340$ & $ 1.17$ & $0.542$ & $1.04$ & $0.860$ & $0.25$ & $0.466$ & $10.0$ & $ 9.5$ & $3.48$ \\ 
$3.88$ & $4$ & $+4$ & $ 5.34$ & $ 0.24$ & $0.256$ & $0.000651$ & $3$ & $-3$ & $ 9.74$ & $2.199$ & $16.9$ & $ 67.02$ & $ 20.2$ & $ 17.7$ & $ 20.0$ & $ 17.3$ & $ 18.0$ & $ 13.5$ & $ 18.3$ & $0.28$ & $1.824$ & $ 1.11$ & $0.547$ & $1.21$ & $0.855$ & $0.27$ & $0.467$ & $10.0$ & $ 9.7$ & $3.46$ \\ 
$3.89$ & $5$ & $+5$ & $ 0.73$ & $ 0.46$ & $0.203$ & $0.000654$ & $4$ & $-1$ & $ 7.47$ & $1.682$ & $40.1$ & $ 63.24$ & $ 20.5$ & $ 19.3$ & $ 20.0$ & $ 20.0$ & $ 19.6$ & $ 13.0$ & $ 20.7$ & $0.05$ & $1.326$ & $ 0.40$ & $0.190$ & $0.52$ & $1.420$ & $0.09$ & $0.269$ & $ 4.0$ & $ 7.2$ & $3.33$ \\ 
$3.92$ & $4$ & $+4$ & $ 0.92$ & $ 0.91$ & $0.256$ & $0.000677$ & $3$ & $+0$ & $ 4.88$ & $2.380$ & $25.9$ & $ 65.68$ & $ 21.3$ & $ 20.2$ & $ 14.9$ & $ 13.3$ & $ 12.5$ & $ 17.4$ & $ 22.6$ & $0.05$ & $1.913$ & $ 0.59$ & $0.206$ & $0.20$ & $2.380$ & $0.07$ & $0.490$ & $10.0$ & $ 6.1$ & $3.87$ \\ 
$3.93$ & $4$ & $+3$ & $ 0.27$ & $ 3.15$ & $0.342$ & $0.000664$ & $3$ & $+1$ & $ 9.84$ & $2.277$ & $56.4$ & $ 64.99$ & $ 21.1$ & $ 17.6$ & $ 19.2$ & $ 15.5$ & $ 16.4$ & $ 14.0$ & $ 21.6$ & $0.04$ & $0.631$ & $ 0.20$ & $0.087$ & $0.39$ & $2.484$ & $0.04$ & $0.215$ & $ 6.2$ & $ 7.2$ & $3.52$ \\ 
$3.93$ & $5$ & $+3$ & $ 0.81$ & $ 0.42$ & $0.339$ & $0.000638$ & $3$ & $-3$ & $ 8.66$ & $2.206$ & $17.0$ & $ 66.28$ & $ 20.8$ & $ 17.8$ & $ 20.6$ & $ 15.6$ & $ 17.3$ & $ 13.6$ & $ 20.2$ & $0.06$ & $1.369$ & $ 1.05$ & $0.532$ & $1.05$ & $0.872$ & $0.27$ & $0.464$ & $ 9.5$ & $ 9.6$ & $3.45$ \\ 
$3.96$ & $5$ & $+4$ & $ 0.24$ & $ 1.85$ & $0.254$ & $0.000638$ & $5$ & $+0$ & $16.91$ & $2.213$ & $68.4$ & $ 59.80$ & $ 20.3$ & $ 18.4$ & $ 20.1$ & $ 18.7$ & $ 16.7$ & $ 13.6$ & $ 20.3$ & $0.03$ & $0.735$ & $ 0.20$ & $0.084$ & $0.76$ & $2.213$ & $0.03$ & $0.187$ & $ 4.2$ & $ 6.8$ & $3.40$ \\ 
$4.03$ & $5$ & $+5$ & $ 0.17$ & $ 2.25$ & $0.203$ & $0.000638$ & $3$ & $+2$ & $ 9.28$ & $2.341$ & $84.4$ & $ 63.07$ & $ 21.6$ & $ 18.2$ & $ 19.4$ & $ 16.9$ & $ 17.7$ & $ 14.5$ & $ 19.0$ & $0.01$ & $1.236$ & $ 0.43$ & $0.089$ & $0.32$ & $2.823$ & $0.02$ & $0.251$ & $ 5.7$ & $ 5.0$ & $3.80$ \\ 
$4.05$ & $5$ & $+4$ & $ 0.65$ & $ 1.29$ & $0.254$ & $0.000638$ & $4$ & $-4$ & $ 2.59$ & $2.158$ & $62.8$ & $ 68.28$ & $ 20.5$ & $ 17.3$ & $ 20.3$ & $ 17.9$ & $ 17.4$ & $ 12.5$ & $ 21.8$ & $0.07$ & $0.870$ & $ 0.58$ & $0.170$ & $0.20$ & $1.297$ & $0.05$ & $0.221$ & $ 4.2$ & $ 6.9$ & $3.39$ \\ 
$4.06$ & $5$ & $+4$ & $ 0.50$ & $ 2.10$ & $0.254$ & $0.000670$ & $3$ & $+1$ & $14.37$ & $2.254$ & $64.4$ & $ 61.06$ & $ 19.7$ & $ 17.5$ & $ 19.7$ & $ 19.1$ & $ 16.0$ & $ 14.4$ & $ 17.8$ & $0.05$ & $0.988$ & $ 0.33$ & $0.101$ & $0.57$ & $2.495$ & $0.03$ & $0.251$ & $ 4.2$ & $ 5.3$ & $3.60$ \\ 
$4.06$ & $4$ & $+4$ & $ 2.87$ & $ 0.53$ & $0.256$ & $0.000669$ & $4$ & $-3$ & $ 0.63$ & $2.220$ & $29.9$ & $ 66.07$ & $ 21.2$ & $ 17.2$ & $ 19.6$ & $ 16.8$ & $ 16.3$ & $ 13.5$ & $ 20.4$ & $0.15$ & $1.902$ & $ 3.04$ & $0.614$ & $0.08$ & $0.794$ & $0.12$ & $0.488$ & $ 4.0$ & $ 5.4$ & $3.58$ \\ 
$4.07$ & $5$ & $+3$ & $ 0.57$ & $ 4.57$ & $0.339$ & $0.000638$ & $3$ & $+2$ & $ 8.39$ & $2.172$ & $66.0$ & $ 54.92$ & $ 21.2$ & $ 17.8$ & $ 18.8$ & $ 13.6$ & $ 13.1$ & $ 11.8$ & $ 17.4$ & $0.10$ & $0.539$ & $ 0.21$ & $0.072$ & $0.33$ & $2.522$ & $0.03$ & $0.183$ & $ 4.9$ & $ 6.5$ & $3.50$ \\ 
$4.11$ & $5$ & $+4$ & $ 0.17$ & $ 2.87$ & $0.254$ & $0.000638$ & $4$ & $+2$ & $ 5.62$ & $2.259$ & $62.7$ & $ 63.56$ & $ 19.1$ & $ 19.2$ & $ 19.9$ & $ 17.5$ & $ 19.2$ & $ 14.0$ & $ 17.8$ & $0.01$ & $1.113$ & $ 0.45$ & $0.101$ & $0.20$ & $2.811$ & $0.02$ & $0.283$ & $ 6.0$ & $ 5.0$ & $3.82$ \\ 
$4.12$ & $3$ & $+3$ & $ 1.51$ & $ 0.54$ & $0.348$ & $0.000656$ & $3$ & $+3$ & $19.14$ & $2.164$ & $14.7$ & $ 64.06$ & $ 18.8$ & $ 18.9$ & $ 18.2$ & $ 16.9$ & $ 16.7$ & $ 13.6$ & $ 18.5$ & $0.05$ & $2.877$ & $ 0.18$ & $0.199$ & $0.37$ & $5.040$ & $0.22$ & $1.001$ & $ 7.5$ & $ 5.0$ & $3.92$ \\ 
$4.13$ & $5$ & $+4$ & $ 0.49$ & $ 2.64$ & $0.254$ & $0.000639$ & $4$ & $+0$ & $25.75$ & $2.782$ & $77.4$ & $ 55.99$ & $ 18.8$ & $ 17.4$ & $ 23.1$ & $ 16.6$ & $ 17.6$ & $ 14.2$ & $ 18.0$ & $0.11$ & $0.431$ & $ 0.06$ & $0.039$ & $0.92$ & $2.782$ & $0.02$ & $0.109$ & $ 7.3$ & $10.4$ & $3.27$ \\ 
$4.13$ & $4$ & $+4$ & $ 0.20$ & $ 2.52$ & $0.256$ & $0.000638$ & $4$ & $+2$ & $ 4.93$ & $2.280$ & $58.7$ & $ 65.09$ & $ 19.2$ & $ 19.8$ & $ 21.3$ & $ 16.2$ & $ 17.6$ & $ 13.7$ & $ 21.7$ & $0.02$ & $1.173$ & $ 0.42$ & $0.105$ & $0.17$ & $2.866$ & $0.03$ & $0.301$ & $ 5.8$ & $ 5.0$ & $3.81$ \\ 
$4.14$ & $5$ & $+4$ & $ 1.39$ & $ 0.45$ & $0.254$ & $0.000639$ & $3$ & $+3$ & $12.69$ & $2.204$ & $18.8$ & $ 64.46$ & $ 18.5$ & $ 18.6$ & $ 18.9$ & $ 16.7$ & $ 16.0$ & $ 12.5$ & $ 19.6$ & $0.04$ & $3.026$ & $ 0.21$ & $0.174$ & $0.28$ & $4.416$ & $0.14$ & $0.769$ & $ 7.6$ & $ 5.1$ & $3.90$ \\ 
$4.17$ & $3$ & $+3$ & $ 0.32$ & $ 3.33$ & $0.348$ & $0.000670$ & $3$ & $-1$ & $ 9.28$ & $2.675$ & $53.3$ & $ 63.98$ & $ 22.3$ & $ 18.6$ & $ 21.6$ & $ 12.9$ & $ 17.0$ & $ 16.5$ & $ 23.9$ & $0.04$ & $0.731$ & $ 0.30$ & $0.104$ & $0.38$ & $2.432$ & $0.04$ & $0.254$ & $ 5.7$ & $ 6.2$ & $3.61$ \\ 
$4.18$ & $4$ & $+3$ & $ 1.42$ & $ 4.58$ & $0.341$ & $0.000673$ & $6$ & $-4$ & $ 2.35$ & $1.869$ & $89.0$ & $ 60.07$ & $ 19.6$ & $ 19.0$ & $ 20.6$ & $ 18.1$ & $ 17.0$ & $ 13.2$ & $ 14.0$ & $0.24$ & $0.566$ & $ 1.20$ & $0.175$ & $0.21$ & $1.106$ & $0.03$ & $0.193$ & $ 4.6$ & $ 6.2$ & $3.52$ \\ 
$4.19$ & $6$ & $+5$ & $ 0.94$ & $ 0.52$ & $0.202$ & $0.000667$ & $4$ & $+1$ & $ 7.41$ & $1.221$ & $49.7$ & $ 63.23$ & $ 19.5$ & $ 19.3$ & $ 18.3$ & $ 17.2$ & $ 16.3$ & $ 13.2$ & $ 18.7$ & $0.09$ & $1.077$ & $ 0.29$ & $0.152$ & $0.51$ & $1.433$ & $0.08$ & $0.218$ & $ 3.4$ & $ 7.5$ & $3.22$ \\ 
\hline
\end{tabular}
\end{center}
{\bf Notes:} The table is sorted in ascending order by the reduced $\chi^2$. For practical reasons, the parameter $a_{\mathrm{sph}}$ is substituted in the fitting procedure with the square root of the mean square of the vertical velocity component when averaged over a spherical surface and oscillation period ${\langle v_\mathrm{v}^2 \rangle}{}^{1/2} = (\pi/2)^{1/2}\nu_\textnormal{osc}\,a_{\mathrm{sph}}$ (cf.~Eq.~(3.136)\cite{2010aste.book.....A}) with $\nu_\textnormal{osc} = \nu_\textnormal{osc}^{(0)} + m \nu_\textnormal{rot} C$ (see Eq.~(\ref{eq:frequency_observed}) and Eq.~(11)\cite{1997A&AS..121..343S}). The quantity $\eta$ is the ratio of the centrifugal force to the gravitational force at the equator of the star (see Eq.~(30)\cite{1997A&AS..121..343S}). The rotation frequency $\nu_\textnormal{rot}$ is deduced from the fitting parameters via Eq.~(\ref{eq:frequency_observed}). The stellar radius $R_\star$ is derived from the identity $v\sin(i) = 2\pi R_\star \nu_\textnormal{rot} \sin(i)$. The stellar mass $M$ (and the surface gravity $g = G M /R_\star^2$) then follows from $M = k^{(0)} (2\pi\nu_\textnormal{osc}^{(0)})^2 R_\star^3 / G$, where $G$ is the gravitational constant (see Eq.~(9)\cite{1997A&AS..121..343S}). Soft limits for $\eta \lesssim 0.25$ and $\nu_\textnormal{rot}/\nu_\textnormal{osc}^{(0)} \lesssim 0.5$ are implemented in the fitting process to remain consistent with the assumptions of the applied slow-rotation model.
\end{sidewaystable*}
\begin{table}
\small 
\caption{Atmospheric parameters, abundances, and photospheric mass fractions derived from spectroscopy.}
\label{table:spectroscopy}
\begin{center}
\begin{tabular}{lr}
\hline\hline
Parameter & Value $\pm$ stat.\ $\pm$ sys.\ \\
\hline
Effective temperature $T_{\mathrm{eff}}$ & $15,570\pm20\pm320$\,K \\
Surface gravity $\log (g\,\mathrm{(cm\,s^{-2})})$ & $3.328\pm0.003\pm0.100$\\
Radial velocity $v_{\mathrm{rad}}$ at MJD\,$54,390.3774$ & $21.3\pm0.2\pm0.2$\,km\,s$^{-1}$ \\
Radial velocity $v_{\mathrm{rad}}$ at MJD\,$54,391.3786$ & $19.4\pm0.2\pm0.1$\,km\,s$^{-1}$ \\
Radial velocity $v_{\mathrm{rad}}$ at MJD\,$54,392.3801$ & $20.5\pm0.2\pm0.2$\,km\,s$^{-1}$ \\
Radial velocity $v_{\mathrm{rad}}$ at MJD\,$55,174.0461$ & $17.0\pm0.2\pm0.2$\,km\,s$^{-1}$ \\
Radial velocity $v_{\mathrm{rad}}$ at MJD\,$55,613.2090$ & $20.1\pm0.2\pm0.3$\,km\,s$^{-1}$ \\
Radial velocity $v_{\mathrm{rad}}$ at MJD\,$55,905.0284$ & $16.7\pm0.3\pm0.3$\,km\,s$^{-1}$ \\
Radial velocity $v_{\mathrm{rad}}$ at MJD\,$56,327.0293$ & $22.0\pm0.3\pm0.4$\,km\,s$^{-1}$ \\
Projected rotational velocity $v\sin(i)$ & $64.2\pm0.1\pm0.3$\,km\,s$^{-1}$ \\
Microturbulence $\xi$ & $1.5\pm0.1\pm1.0$\,km\,s$^{-1}$ \\
Helium abundance $\log(n(\textnormal{He}))$ & $-0.860\pm0.003\pm0.109$ \\
Carbon abundance $\log(n(\textnormal{C}))$ & $-5.462\pm0.021\pm0.050$ \\
Nitrogen abundance $\log(n(\textnormal{N}))$ & $-3.239\pm0.006\pm0.086$ \\
Oxygen abundance $\log(n(\textnormal{O}))$ & $-3.269\pm0.008\pm0.060$ \\
Neon abundance $\log(n(\textnormal{Ne}))$ & $-4.108\pm0.004\pm0.012$ \\
Magnesium abundance $\log(n(\textnormal{Mg}))$ & $-4.611\pm0.011\pm0.121$ \\
Aluminum abundance $\log(n(\textnormal{Al}))$ & $-5.729\pm0.010\pm0.025$ \\
Silicon abundance $\log(n(\textnormal{Si}))$ & $-4.442\pm0.013\pm0.259$ \\
Sulfur abundance $\log(n(\textnormal{S}))$ & $-4.886\pm0.005\pm0.020$ \\
Argon abundance $\log(n(\textnormal{Ar}))$ & $-5.525\pm0.017\pm0.076$ \\
Iron abundance $\log(n(\textnormal{Fe}))$ & $-4.601\pm0.007\pm0.144$ \\
\hline\hline
Derived quantity & Value $\pm$ sys. \\
\hline
Hydrogen mass fraction $X$ & $0.602\pm0.062$ \\
Helium mass fraction $Y$ & $0.383\pm0.062$ \\
Metallicity $Z \equiv 1-X-Y$ & $0.015\pm0.001$ \\
\hline
\end{tabular}
\end{center}
{\bf Notes:} The abundance $n(x)$ is given as fractional particle number of species $x$ with respect to all elements. Statistical uncertainties (\textit{``stat.''}) correspond to $\Delta \chi^2 = 6.63$ and are 99\% confidence limits. Systematic uncertainties (\textit{``sys.''}) cover only the effects induced by additional variations of $2\%$ in $T_{\textnormal{eff}}$ and $0.1$ in $\log(g)$ and are formally taken to be 99\% confidence limits\cite{2014A&A...565A..63I}.
\end{table}
\end{document}